\documentclass[aps,prl,reprint,superscriptaddress,longbibliography,floatfix]{revtex4-2}

\usepackage{mathtools,amssymb,graphicx}
\usepackage{amsmath,bm}
\usepackage[colorlinks=true,linkcolor=blue,urlcolor=blue,citecolor=blue,
pdftitle={Field-amplified readouts of weak altermagnetic exchange in MnF2}]{hyperref}

\newcommand{\mnf}{$\mathrm{MnF_2}$}
\newcommand{\fef}{$\mathrm{FeF_2}$}
\newcommand{\Q}{\mathbf Q}
\newcommand{\kvec}{\mathbf k}
\newcommand{\dJ}{\delta J_7}
\newcommand{\lam}{\lambda_{\rm AM}}
\DeclareMathOperator*{\argmin}{arg\,min}

\begin{document}

\title{Field-amplified readouts of weak altermagnetic exchange in \mnf{}}

\author{Guowen Jiang}
\affiliation{Lanzhou Center for Theoretical Physics, Lanzhou University, Lanzhou 730000, China}
\affiliation{Key Laboratory of Quantum Theory and Applications of MoE, Lanzhou University, Lanzhou 730000, China}
\affiliation{Key Laboratory of Theoretical Physics of Gansu Province and Gansu Provincial Research Center for Basic Disciplines of Quantum Physics, Lanzhou University, Lanzhou 730000, China}

\author{Feilong Wang}
\affiliation{Lanzhou Center for Theoretical Physics, Lanzhou University, Lanzhou 730000, China}
\affiliation{Key Laboratory of Quantum Theory and Applications of MoE, Lanzhou University, Lanzhou 730000, China}
\affiliation{Key Laboratory of Theoretical Physics of Gansu Province and Gansu Provincial Research Center for Basic Disciplines of Quantum Physics, Lanzhou University, Lanzhou 730000, China}

\author{Yunhua Wang}
\affiliation{Lanzhou Center for Theoretical Physics, Lanzhou University, Lanzhou 730000, China}
\affiliation{Key Laboratory of Quantum Theory and Applications of MoE, Lanzhou University, Lanzhou 730000, China}
\affiliation{Key Laboratory of Theoretical Physics of Gansu Province and Gansu Provincial Research Center for Basic Disciplines of Quantum Physics, Lanzhou University, Lanzhou 730000, China}

\author{Fawei Zheng}
\affiliation{Centre for Quantum Physics, Key Laboratory of Advanced Optoelectronic Quantum Architecture and Measurement (MOE), School of Physics, Beijing Institute of Technology, Beijing 100081, China}
\affiliation{Beijing Key Laboratory of Nanophotonics and Ultrafine Optoelectronic Systems, School of Physics, Beijing Institute of Technology, Beijing 100081, China}
\affiliation{International Center for Quantum Materials, Beijing Institute of Technology, Zhuhai 519000, China}

\author{Bin Xi}
\email{binxi@yzu.edu.cn}
\affiliation{College of Physics Science and Technology, Yangzhou University, Yangzhou 225002, China}

\author{Hong-Gang Luo}
\affiliation{Lanzhou Center for Theoretical Physics, Lanzhou University, Lanzhou 730000, China}
\affiliation{Key Laboratory of Quantum Theory and Applications of MoE, Lanzhou University, Lanzhou 730000, China}
\affiliation{Key Laboratory of Theoretical Physics of Gansu Province and Gansu Provincial Research Center for Basic Disciplines of Quantum Physics, Lanzhou University, Lanzhou 730000, China}

\author{Jize Zhao}
\email{zhaojz@lzu.edu.cn}
\affiliation{Lanzhou Center for Theoretical Physics, Lanzhou University, Lanzhou 730000, China}
\affiliation{Key Laboratory of Quantum Theory and Applications of MoE, Lanzhou University, Lanzhou 730000, China}
\affiliation{Key Laboratory of Theoretical Physics of Gansu Province and Gansu Provincial Research Center for Basic Disciplines of Quantum Physics, Lanzhou University, Lanzhou 730000, China}
\begin{abstract}
\mnf{}, the textbook two-sublattice antiferromagnet, has reemerged as a prototypical altermagnet, yet the sublattice-odd exchange that defines
this identity remains under active debate: it enters the magnon splitting only in quadrature with the dipole--dipole interaction, its magnitude
suppressed and its sign erased. An overdetermined first-principles total-energy mapping resolves this scale as a seventh-neighbor imbalance
$\dJ\simeq+8~\mu$eV. The resulting Hamiltonian, with the dipole--dipole interaction included explicitly, reproduces the
low-energy gap and the visible finite-momentum splitting. A longitudinal field $B\parallel c$ then acts as a linear amplifier of the hidden scale,
opening two signed, field-linear readouts. The first is the compensation field $B^\ast(\Q)$, the position of minimum splitting, which is equal and
opposite at the rotation-related partner momenta: a shift from zero field is itself evidence of a finite imbalance, its side gives the sign, and its
magnitude, $|B^\ast|\simeq0.34$~T here, gives the scale. The second is the fixed-field contrast of the partner splittings, $\simeq0.14$~meV at
$1$~T, six times the zero-field excess: a sign check from just two spectra. Both readouts survive a $0.12$~meV energy resolution, and the
construction carries over to any easy-axis collinear altermagnet, bringing $\mu$eV altermagnetic exchange within present instrumental reach.
\end{abstract}

\maketitle

An altermagnet is a collinear magnet with vanishing net magnetization in which the opposite-spin sublattices are connected by a symmetry operation
containing a crystal rotation, rather than by a lattice translation or inversion~\cite{amintroPhysRevX.12.031042,amintroPhysRevX.12.040002,amintroPhysRevX.12.040501}.
This rotation permits a momentum-dependent spin splitting that is forbidden in an ordinary N\'eel antiferromagnet and that reverses sign
between rotation-related momenta. The magnons inherit this pattern as chiral modes of alternating handedness~\cite{chiral-magnon-PhysRevLett.131.256703,chiral-magnon-PhysRevLett.133.156702}, to which polarized neutron
scattering couples directly~\cite{pins-PhysRevB.111.L060405}. \mnf{}, the textbook two-sublattice N\'eel antiferromagnet, possesses precisely this
rutile magnetic symmetry: a giant momentum-dependent spin splitting was predicted here even before altermagnetism was
formalized~\cite{PhysRevB.102.014422}, and first-principles work has identified the correlated $d$-wave splittings of its electronic and magnon bands~\cite{mnf-25ph-c3kh-5s5t}. 

Yet the scale behind these predictions has resisted zero-field measurement. Unpolarized inelastic neutron scattering (INS) resolves no
altermagnetic band splitting within instrumental resolution~\cite{magnon-absencePhysRevLett.134.226702}. Polarized INS
(PINS)~\cite{pins-faure2025altermagnetismrevealedpolarizedneutrons}, in contrast, resolves a magnon doublet, attributes the visible splitting mainly to the long-range dipole--dipole interaction (DDI), and
detects opposite chiral responses in domain-biased samples.
Microscopic electronic-structure analysis finds the altermagnetic parameters small~\cite{mnf-band-solovyev2026altermagnetismmnf2bandsplitting},
complemented by diffraction and multipole views of the order~\cite{lovesey2026chiralorder, multipolar-PRL-2026}. The existing determinations of the
sublattice-odd exchange asymmetry---the weak scale that carries this altermagnetism---differ even in the sign of the central value
(SM~\cite{SMnote} Sec.~S9), and the deadlock is one of principle rather than of instrumentation: the weak exchange enters the zero-field splitting only
in quadrature with the dominant DDI, and better resolution cannot restore a sign the spectrum never contains. What is missing is not a sharper
zero-field spectrum but a different observable that is linear in the weak exchange and carries its sign.

Constructing that observable requires both competing scales in hand, and first-principles theory supplies them. Effective spin models assign the
zone-center gap of about $1$~meV to an axial single-ion anisotropy $D_c$~\cite{magnon-absencePhysRevLett.134.226702, rezende2019afmmagnons}, yet high-spin
$3d^5$ Mn$^{2+}$ (spin $S=5/2$, orbital moment nearly quenched) supports only a weak relativistic single-ion term, and the long-range DDI is the
leading microscopic source of the anisotropy and the gap~\cite{ddi-Keffer1952MnF2}. An overdetermined mapping of
first-principles total energies (End Matter, Appendix~A) resolves the individual exchange paths and yields a weak but finite seventh-neighbor
imbalance $\dJ=J_{7b}-J_{7a}\simeq+7.96~\mu$eV, while spin--orbit calculations confirm the small single-ion anisotropy (SM~\cite{SMnote} Sec.~S3). 
With the Ewald-summed DDI and no fitted anisotropy, linear spin-wave theory reproduces the measured gap and the experimental-scale
finite-momentum splitting. Both competing scales are thus fixed: a dipolar mixing that dominates
the visible fine structure, and a smaller sublattice-odd detuning of the same doublet, masked by the former.

How can that smaller detuning be read out? Let $\Omega_{\Q}(B)$ be the splitting of the target magnon doublet under a magnetic field $B$ at momentum $\Q$ (End Matter; SM~\cite{SMnote} Sec.~S5), 
an experimentally accessible quantity. The weak exchange enters it as a signed intrinsic detuning $\Lambda_{\Q}$, odd between partner momenta with amplitude $4S|\dJ|$; 
the DDI enters as a transverse mixing whose projected strength $D_{\Q}$ is common to the pair. The two combine in quadrature, so at zero field the detuning 
contributes only an excess splitting $\delta\Omega(0)$ that is quadratic in $\Lambda_{\Q}$ and blind to its sign. A longitudinal field $B\parallel c$ removes 
both obstacles: it adds a linear Zeeman detuning, and the splitting is minimal where the two detunings cancel, at the momentum-resolved compensation field $B^\ast(\Q)=-\Lambda_{\Q}/2g_{\rm eff}\mu_B$, with $g_{\rm eff}$ the effective Zeeman factor of the doublet and $\mu_B$ the Bohr magneton~(SM~\cite{SMnote} Sec.~S2 and S5). 
A shift of the minimum from zero field is itself evidence of a finite detuning, its side gives the sign, its size gives $\lvert\Lambda_{\Q}\rvert$. The same shift renders the splitting asymmetric under $B\to-B$; the resulting partner contrast at 
one fixed field is a rapid sign check that no single spectrum can provide. For $\Q_A=(0.25,0.75,0)$ and $\Q_{A'}=C_{4z}\Q_{A}$, with $C_{4z}$ the fourfold rotation about the $c$ axis, the present parameters give $B^\ast_{A}=-B^\ast_{A'}\simeq\mp0.344$~T (SM~\cite{SMnote} Sec.~S2); the shift survives a $0.12$~meV energy resolution~(SM~\cite{SMnote} Sec.~S7).
This is precisely the field amplification announced in the title,
\begin{equation}
\delta\Omega(0)\propto\Lambda_{\Q}^2 \quad\Longrightarrow\quad B^\ast(\Q)\propto\Lambda_{\Q}:
\label{eq:amplification}
\end{equation}
the weak scale is read no longer as a tiny quadratic energy correction but as a linear, signed position on the field axis.

\textit{Microscopic model and first-principles exchange.---}The model contains isotropic Heisenberg exchange between the Mn spins, the long-range DDI, and the Zeeman coupling to the applied field; the full Hamiltonian, its symbols, and its sign conventions are collected in the End Matter, Eqs.~\eqref{eq:H}--\eqref{eq:HZ}, with the exchange written so that a positive $J_{ij}$ is antiferromagnetic.

\begin{figure}[!t]
\centering
\includegraphics[width=0.90\columnwidth]{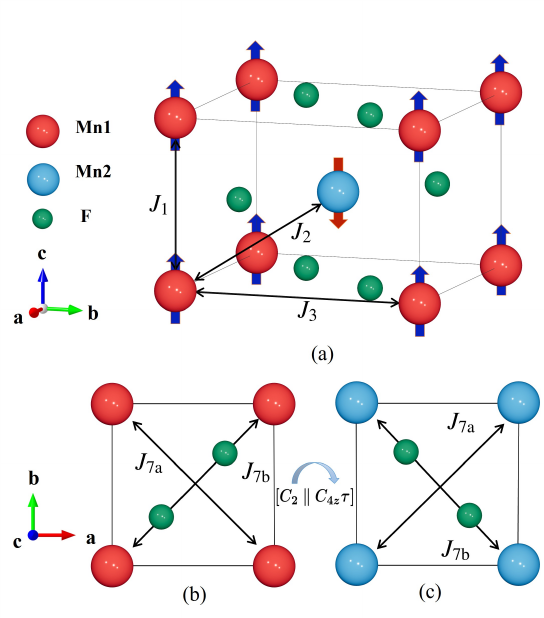}
	\caption{Structure and exchange paths of \mnf{}. (a) Rutile magnetic cell with the two antiparallel Mn sublattices (Mn1 red, Mn2 blue), the F ions (green), and the near-neighbor exchange 
paths $J_1$, $J_2$, and $J_3$; bold arrows indicate the antiparallel moments along the $c$ axis. (b),(c) $ab$-plane projection of the two seventh-neighbor paths as seen from sublattice Mn1 (b) and Mn2 (c). 
$J_{7a}$ and $J_{7b}$ have equal length and are interchanged by the spin-group operation $[C_2\,\|\,C_{4z}\bm \tau]$ connecting the sublattices---hence the exchanged labels between the panels---but 
only $J_{7b}$ is bridged by two fluorine ions.}
\label{fig:lattice}
\end{figure}

In the rutile structure of \mnf{} the Mn ions at the cell corner and at the body center form the two magnetic sublattices, ordered antiparallel along $c$ below $T_N\approx67$~K~\cite{mnf-neutron-doi:10.1139/P10-081,mnf-neutron-osti_4092064}. No translation or inversion maps one sublattice onto the other; the connecting operation involves the fourfold screw of the lattice, and its first consequence for the exchange network appears in the seventh-neighbor shell [Fig.~\ref{fig:lattice}]: two bond families of identical length, $J_{7a}$ and $J_{7b}$, interchanged by the screw---symmetry equivalent as a pair, crystallographically distinct individually, only the $J_{7b}$ path bridged by two fluorine ions. The imbalance $\dJ=J_{7b}-J_{7a}$ is therefore the leading sublattice-odd exchange scale of \mnf{} and the microscopic carrier of its altermagnetism.

The couplings were determined from total energies alone: a least-squares inversion of the overdetermined configuration--energy system---several hundred collinear magnetic configurations in two optimized supercells, with no neutron input (End Matter, Appendix~A)---separates the individual exchange paths and gives $\dJ\simeq7.96~\mu$eV from the $28$-Mn supercell and $7.79~\mu$eV from the $20$-Mn one, consistent to $2.1\%$; the former is used throughout. Bootstrap statistics, the complete coupling tables, the $U_{\rm eff}$ dependence, and convergence tests are collected in the Supplemental Material (SM~\cite{SMnote}).

With the microscopic single-ion anisotropy confirmed small (SM~\cite{SMnote} Sec.~S3), we retain the DDI explicitly instead of a fitted $D_c$,
evaluating its long-range lattice sum by Ewald summation~\cite{ewald-PhysRevB.70.174426}. Linear spin-wave theory~\cite{lswt-PhysRev.58.1098,lswt-colpa-COLPA1978327,lswt-Toth_2015} on the mapped Hamiltonian 
establishes three facts on which this Letter builds: the DDI by itself opens a $\Gamma$-point gap of $1.15$~meV, matching the measured scale of about $1.08$~meV~\cite{mnf-neutron-doi:10.1139/P10-081,mnf-neutron-osti_4092064};
at generic momenta the DDI produces the finite-momentum hybridization and doublet splitting that dominate the visible fine structure---at $\Q=(\tfrac12,\tfrac12,1)$, where the maximal splitting $178\pm3~\mu$eV is reported~\cite{pins-faure2025altermagnetismrevealedpolarizedneutrons}, the calculated $179~\mu$eV is entirely dipolar, the imbalance form factor vanishing there as at $\Gamma$ [Eq.~\eqref{eq:lam_rlu} below; 
SM~\cite{SMnote}]; and $\dJ$ contributes a smaller sublattice-odd detuning of the same doublet, $|\Lambda_{\Q}|=79.6~\mu$eV against the projected dipolar mixing $D_{\Q}=121.6~\mu$eV at 
the working momenta of this Letter.
An effective $D_c$ may remain a valid low-energy parametrization, but it compresses two physically different contributions---the nonlocal DDI, whose doublet projection is the momentum-dependent mixing $D_{\Q}$ used below, and the weak spin--orbit single-ion term---into one local constant, hiding exactly the momentum structure the readouts exploit.

\textit{Sublattice-odd magnon detuning.---}The exchange couplings within the two magnetic sublattices define a sublattice-odd form factor $\lam(\kvec)$, which reverses sign under any space-group operation $g$ that interchanges the sublattices, $\lam(g\kvec)=-\lam(\kvec)$ (End Matter). For the seventh-neighbor imbalance of \mnf{}, with $\kvec=(h,k,l)$ in reciprocal-lattice units (r.l.u.),
\begin{equation}
\lam(h,k,l)=4S\,\dJ\,\sin(2\pi h)\sin(2\pi k).
\label{eq:lam_rlu}
\end{equation}
We work at the partner momenta $\Q_A=(0.25,0.75,0)$ and $\Q_{A'}=(-0.75,0.25,0)=C_{4z}\Q_A$, labeled $q=A,A'$. Because the ordering of the two tracked modes is a convention of its own, we keep $\lam$ for the exchange form factor and write $\Lambda_q$ for the signed intrinsic detuning of the tracked doublet; in the mode-ordering convention of this work (SM~\cite{SMnote}),
\begin{equation}
\Lambda_q=-\lam(\Q_q),
\label{eq:Lambda_conv}
\end{equation}
so that $\dJ>0$ implies $\Lambda_{A}=+4S\dJ=-\Lambda_{A'}$.

The opposite detunings are protected by an exact symmetry of the full problem. The magnetic space group of \mnf{} contains the antiunitary element $\{C_{4z}|\bm\tau\}\mathcal T$ combining the fourfold 
screw with time reversal $\mathcal T$: the screw interchanges the sublattices, $\mathcal T$ restores the N\'eel domain, and $\mathcal T$ simultaneously reverses the applied field. 
Acting on the magnon problem, this element carries $(\Q_A,B)$ into $(\Q_{A'},-B)$, so the full bosonic Bogoliubov--de Gennes (BdG) spectrum, dipolar term included, satisfies
\begin{equation}
\Omega_{A'}(B)=\Omega_{A}(-B),
\label{eq:mirror}
\end{equation}
with $\Omega_{A}$, $\Omega_{A'}$ the tracked doublet splittings at the two momenta for $B\parallel c$---exact within the model of Eqs.~\eqref{eq:Hex}--\eqref{eq:HZ} (transformation in the SM~\cite{SMnote}).

\textit{Field-amplified readouts.---} We now formalize the quantities used in the introduction. The tracked doublet has branches $\omega_\pm(\Q,B)$ and splitting $\Omega_{\Q}(B)=\omega_+(\Q,B)-\omega_-(\Q,B)$. Projected onto this two-mode subspace (End Matter), the quadratic spin-wave problem carries the diagonal detuning $\Lambda_{\Q}+h_{\Q}(B)$, with $h_{\Q}(B)$ the field-induced part, together with transverse mixing amplitudes $d_{x,y}(\Q)$ obtained by projecting the Ewald-summed dipolar tensor onto the doublet; their combined scale is $D_{\Q}=2\sqrt{d_x^2(\Q)+d_y^2(\Q)}$, which is a magnon-mode mixing energy. For $B\parallel c$, well below the $9.3$~T spin-flop field~\cite{Jacobs1961spinflop}, the two branches shift linearly and in opposite senses, so $h_{\Q}(B)\simeq2g_{\rm eff}\mu_BB$ with $2g_{\rm eff}\mu_B=231.5~\mu$eV/T for $g_{\rm eff}\simeq2$, and the doublet splitting obeys
\begin{equation}
\Omega^2_{\Q}(B)=\bigl[\Lambda_{\Q}+2g_{\rm eff}\mu_BB\bigr]^2+D^2_{\Q},
\label{eq:master}
\end{equation}
in which the three energy scales stay separated. For the partner pair the full symmetry gives $\Lambda_{A'}=-\Lambda_{A}$ and $D^2(\Q_{A})=D^2(\Q_{A'})$.

At fixed momentum, abbreviate $\Omega\equiv\Omega_{\Q}$, $\Lambda\equiv\Lambda_{\Q}$, $D\equiv D_{\Q}$. Zero field gives $\Omega(0)=\sqrt{D^2+\Lambda^2}$; for $|\Lambda|\ll D$ the excess over the $\Lambda=0$ reference is $\delta\Omega(0)\equiv\Omega(0)-D\simeq\Lambda^2/2D$, quadratic and even in $\Lambda$---the expansion serves only to exhibit this quadratic, sign-blind response; all \mnf{} numbers below are evaluated with the exact expression. The \mnf{} numbers, at $\Q_A$ and equally at $\Q_{A'}$, show what this costs: $D=121.6~\mu$eV and $|\Lambda|=79.6~\mu$eV give $\Omega(0)=145.3~\mu$eV, so the altermagnetic information in a zero-field measurement is a $24~\mu$eV excess on a dipolar baseline that must itself be known, independently, to a few $\mu$eV. The problem is not that the altermagnetic scale is negligible---$|\Lambda|/D\simeq0.65$, the same order of magnitude---but that the two scales add in quadrature: at zero field $\Lambda$ enters Eq.~\eqref{eq:master} only as $\Lambda^2$ under the square root, which suppresses its contribution and erases its sign. 

\begin{figure}[!t]
\centering
\includegraphics[width=\columnwidth]{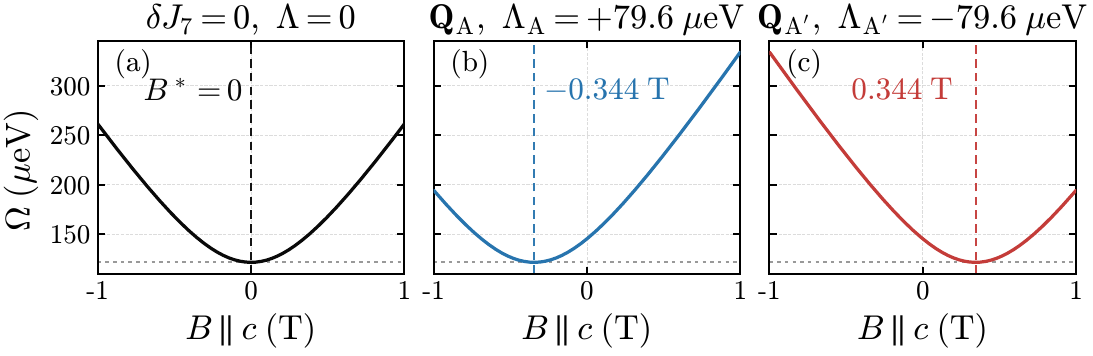}
\caption{Two-mode compensation mechanism, Eq.~\eqref{eq:master}, for fixed $D=121.6~\mu$eV and $2g_{\rm eff}\mu_B=231.5~\mu$eV/T. (a) $\dJ=0$ ($\Lambda=0$), (b) $\Q_A$, $\Lambda_A=+79.6~\mu$eV, (c) $\Q_{A'}$, $\Lambda_{A'}=-79.6~\mu$eV. In all panels the minimum floor equals $D$; the detuning only translates the minimum, to $B^\ast=0$, $-0.344$~T, and $+0.344$~T (dashed lines). 
Panel (a) is the null case, the minimum pinned at $B=0$; panels (b) and (c) realize the two partner momenta of \mnf{} at a common field, 
and their fixed-field difference is the directly measurable sign contrast.}
\label{fig:twomode}
\end{figure}

A finite field escapes this. The minimum of $\Omega_{\Q}(B)$ obeys $\partial\Omega_{\Q}/\partial B=0$, which by Eq.~\eqref{eq:master} is the linear condition $\Lambda_{\Q}+2g_{\rm eff}\mu_BB^\ast(\Q)=0$, defining the momentum-resolved compensation field
\begin{equation}
B^\ast(\Q)\equiv\argmin_{B}\,\Omega_{\Q}(B)=-\frac{\Lambda_{\Q}}{2g_{\rm eff}\mu_B}\propto\Lambda_{\Q}.
\label{eq:Bstar}
\end{equation}
From $\Lambda(g\Q)=-\Lambda(\Q)$ follows $B^\ast(g\Q)=-B^\ast(\Q)$ and thus $B^\ast_{A}=-B^\ast_{A'}$---the low-field, two-mode content of the exact mirror relation~\eqref{eq:mirror}. Figure~\ref{fig:twomode} shows the mechanism at the \mnf{} scales: one and the same $D$ with $\Lambda=0,\pm79.6~\mu$eV shifts the minimum of $\Omega(B)$ to $B^\ast=0,\mp0.344$~T while its floor stays put. For $\Lambda=0$ the splitting is even in $B$ and its minimum is pinned at $B=0$ [Fig.~\ref{fig:twomode}(a)]; a minimum displaced from zero field is therefore, by itself, evidence of a finite sublattice-odd detuning, and the side to which it is displaced is the sign of $\Lambda$. The same sign is readable without locating the minimum, though never from one spectrum---a single absolute splitting fixes no sign: the displacement makes $\Omega(B)$ asymmetric under $B\to-B$, so that within one experiment $\Omega(+1~\mathrm{T})$ and $\Omega(-1~\mathrm{T})$---equivalently, by the mirror relation~\eqref{eq:mirror}, the two partner splittings at a common field of $1$~T---differ by $\simeq0.14$~meV [Figs.~\ref{fig:twomode}(b),(c)]---six times the $24~\mu$eV zero-field excess---an asymmetry that Eq.~\eqref{eq:C} below turns into a differential observable.

\textit{Field-dependent neutron spectra in \mnf.---}We evaluate the full BdG problem with the mapped exchange couplings, the Ewald-summed DDI, and $\mathbf B\parallel c$. With $\dJ=7.96~\mu$eV and $S=5/2$, $|\Lambda_{A}|=4S|\dJ|=79.6~\mu$eV, and Eq.~\eqref{eq:Bstar} places the minima at
\begin{equation}
B^\ast_{A}=-0.344~\text{T},\qquad B^\ast_{A'}=+0.344~\text{T},
\label{eq:BstarAB}
\end{equation}
the overall signs fixed by the bond, momentum, and domain conventions documented in the SM~\cite{SMnote}. A full-BdG scan over several $C_{4z}$-related partner pairs confirms that neither relation is tied to this working point: $B^\ast$ follows the sublattice-odd form factor continuously across the family, reverses sign within every pair, and matches Eq.~\eqref{eq:Bstar} to numerical accuracy (SM~\cite{SMnote} Sec.~S5.7).

The equal-$J_7$ reference ($\dJ=0$, identical DDI) retains every generic feature: its zero-field gap persists, its finite-momentum dipolar splitting stays fully visible, and the field still moves the magnon energies---the doublet broadens from $122$ to $262~\mu$eV at $1$~T. The one response it cannot produce is a displaced minimum: with no sublattice-odd detuning the splitting is even in $B$, and the minima stay locked at $B=0$.
\begin{figure}[!t]
\centering
\includegraphics[width=\columnwidth]{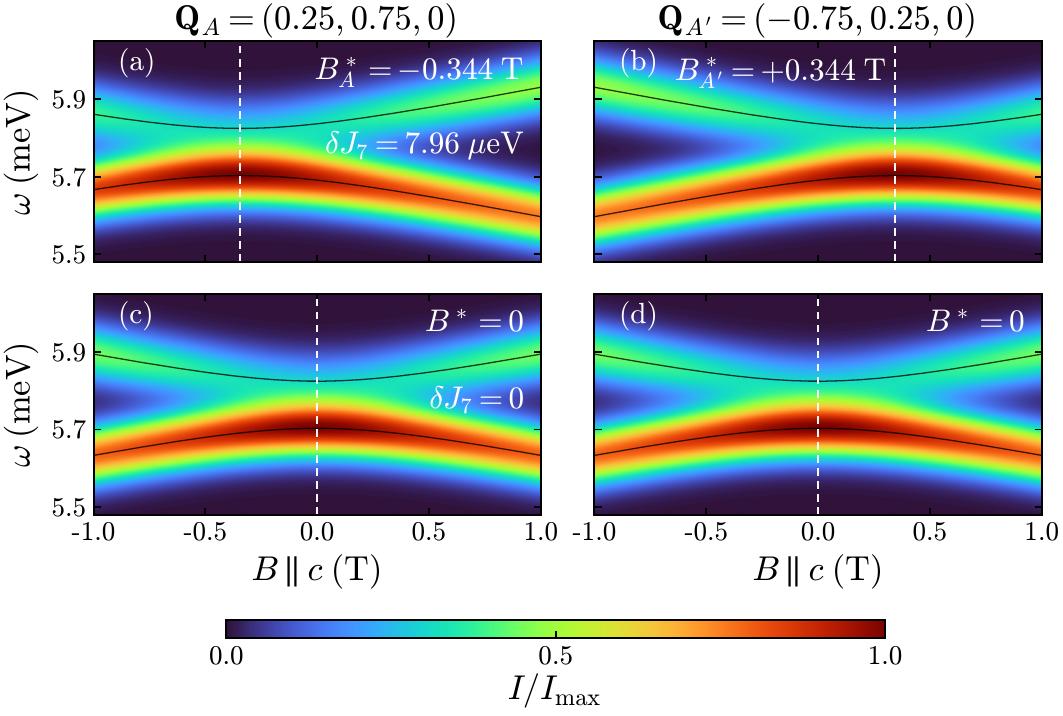}
\caption{Resolution-convolved unpolarized INS intensity $I/I_{\max}$ versus energy transfer $\omega$ and field $B\parallel c$ at $\Q_{A}=(0.25,0.75,0)$ (left column) and $\Q_{A'}=(-0.75,0.25,0)$ (right column), from the full BdG calculation with Ewald-summed DDI and Gaussian energy convolution of $0.12$~meV FWHM; black lines mark the unconvolved branch energies. (a),(b) Physical model, $\dJ=7.96~\mu$eV: the doublet is narrowest at $B^\ast_{A}=-0.344$~T and $B^\ast_{A'}=+0.344$~T (dashed lines), the two maps being mirror images about $B=0$ as Eq.~\eqref{eq:mirror} requires. (c),(d) Equal-$J_7$ reference, $\dJ=0$ with identical DDI: the dipolar splitting and its field dependence remain, but both minima are pinned at $B^\ast=0$ and each map is symmetric in $B$.}
\label{fig:spectra}
\end{figure}

Figure~\ref{fig:spectra} presents the resulting unpolarized INS intensity at both momenta versus energy and field, convolved with a Gaussian energy resolution of $0.12$~meV 
full width at half maximum (FWHM) at a target energy near $5.8$~meV, slightly broader than the $\simeq0.10$~meV resolution achieved in the \mnf{} HYSPEC experiment of Ref.~\cite{pins-faure2025altermagnetismrevealedpolarizedneutrons}. Convolution does not endanger the signature, because the signature is a position along the field axis rather than an energy difference: resolution smears the doublet in $\omega$ but does not move the field of closest approach. Nor need the $24~\mu$eV depth of the valley be resolved at any single field: $B^\ast$ is fixed by the curvature of the entire scan, and global fits of synthetic counting data at $0.10$--$0.20$~meV FWHM recover $\dJ$ to within $6\%$ (SM~\cite{SMnote}). The floor carries independent information as well: Eq.~\eqref{eq:master} gives $\Omega=D_{\Q}$ at $B^\ast$, so a single field scan separates the sublattice-odd diagonal detuning from the projected transverse mixing scale.

\textit{Quantitative extraction and discussion.---}The magnitude of the compensation field converts directly into the weak scale: $|\Lambda_{A}|=2g_{\rm eff}\mu_B|B^\ast_{A}|$, and within the \mnf{} microscopic model, where $|\Lambda_{A}|=4S|\dJ|$,
\begin{equation}
|\dJ|=\frac{g_{\rm eff}\mu_B|B^\ast_{A}|}{2S}.
\label{eq:extraction}
\end{equation}
With $S=5/2$ fixed by the $3d^5$ configuration and $g_{\rm eff}$ independently known for \mnf{}, no additional fitting parameter enters this conversion.

\begin{figure}[!t]
\centering
\includegraphics[width=\columnwidth]{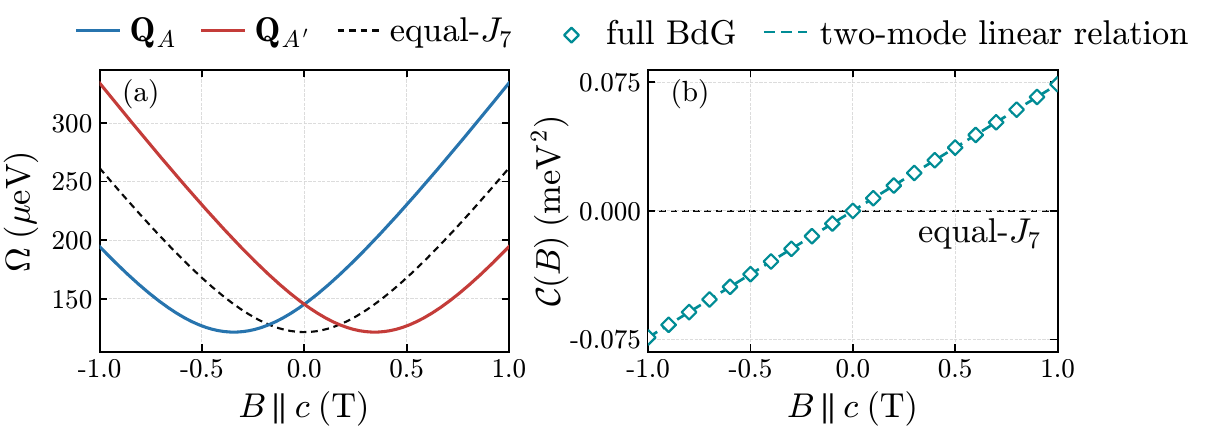}
\caption{Field dependence of the magnon splitting and its sublattice-odd combination. (a) Full BdG splittings $\Omega_{A}(B)$ and $\Omega_{A'}(B)$ at $\Q_{A}$ and $\Q_{A'}$ for finite $\dJ$, together with the equal-$J_7$ reference. The finite exchange imbalance shifts the two minima to opposite fields, while the reference remains centered at zero field; all three curves share the minimum floor $D$. (b) $\mathcal C(B)=\Omega_{A}^2(B)-\Omega_{A'}^2(B)$: open diamonds, full BdG; dashed line, the two-mode relation~\eqref{eq:C} with slope $0.074~\mathrm{meV^2/T}$. The equal-$J_7$ reference remains zero within numerical accuracy.}
\label{fig:CofB}
\end{figure}

The partner pair casts the fixed-field sign check as a differential observable, the squared-splitting difference
\begin{equation}
\mathcal C(B)=\Omega^2_{A}(B)-\Omega^2_{A'}(B).
\label{eq:C_def}
\end{equation}
With $\Lambda_{A'}=-\Lambda_{A}$ and $D^2(\Q_{A})=D^2(\Q_{A'})$, the common projected mixing term $D^2_{\Q}$ in Eq.~\eqref{eq:C_def} cancels identically and leaves the two-mode relation
\begin{equation}
\mathcal C(B)=8g_{\rm eff}\mu_B\Lambda_{A}B,
\qquad \mathcal C(-B)=-\mathcal C(B),
\label{eq:C}
\end{equation}
odd in the field and linear in the weak scale. The full BdG calculation follows this closely [Fig.~\ref{fig:CofB}]: the computed response is odd and nearly linear over $\pm1$~T with slope $0.074~\mathrm{meV^2/T}$ (an $\Omega_{A}-\Omega_{A'}$ difference of $\simeq0.14$~meV at $1$~T), and the equal-$J_7$ reference yields $\mathcal C(B)\simeq0$ to numerical accuracy. The two readouts are complementary in their demands and systematics: $B^\ast$ uses a single momentum but a full field scan, trading an energy discrimination for a positional one---the instrument locates a minimum displaced by $0.34$~T instead of resolving a $24~\mu$eV excess---whereas $\mathcal C(B)$ uses the momentum pair at one fixed field, where the mixing term has already cancelled and a common drift of the doublet center drops out. The equal-$J_7$ reference collapses both to sharp nulls, so agreement of the two readouts on a single $\Lambda_A$ is an internal consistency test of the two-mode picture, not a redundancy.

Three remarks bound the scope of these results. First, beyond the two readouts, a global fit of all spectra---every field and both momenta in one likelihood, all effective parameters 
free---refines the extraction further (SM~\cite{SMnote}). Second, Eq.~\eqref{eq:master} is a low-energy effective relation; the statement that holds exactly is the mirror relation~\eqref{eq:mirror}. 
Third, a N\'eel-domain fraction $p$ dilutes the odd response by $2p-1$, so $B^\ast(\Q)$ is defined relative to the majority domain, fixed by the domain-biased configuration demonstrated in Ref.~\cite{pins-faure2025altermagnetismrevealedpolarizedneutrons}.

\textit{Discussion.---} Polarized neutrons measure magnon chirality and domain populations directly; that is how the \mnf{} doublet was resolved and its branches identified as 
counter-rotating~\cite{pins-PhysRevB.111.L060405,pins-faure2025altermagnetismrevealedpolarizedneutrons}. The finite-field protocol reads the complementary projection---the sublattice-odd energy scale between 
the modes whose chirality and eigenvector content polarized INS returns; combined, the two overdetermine the low-energy Hamiltonian. High-resolution neutron work on \fef{} reveals the same coexistence 
of dominant dipolar and weaker altermagnetic exchange splittings~\cite{pins-g6dt-rf8c}; the scheme carries over there---and across 
the rutile fluoride family---with correspondingly smaller compensation fields (SM~\cite{SMnote} Sec. S8). 
Finite-field neutron studies of MnTe use the field to select or switch domains and their chirality~\cite{pins-m8lc-f8gk} or to rotate the N\'eel vector~\cite{magnon-neelfield2026}: there the field 
manipulates the configuration; here the configuration is held fixed and the field is an energy detuning, reading out a weak intrinsic magnon scale of the ordered state.

\textit{Conclusion.---}A weak sublattice-odd scale need not be read as a small energy; it can be read as a position. In \mnf{}, the mapped seventh-neighbor imbalance---$\dJ\simeq8~\mu$eV, a few percent of the dominant exchange---enters the zero-field splitting only quadratically, beneath the dipolar fine structure; a longitudinal field converts it into signed positions on the field axis, the compensation minima 
at $B^\ast_{A}=-B^\ast_{A'}\simeq\mp0.34$~T---the displacement robust at $0.12$~meV resolution---and into the $0.14$~meV fixed-field partner contrast that two spectra deliver. Nothing in the construction 
is specific to \mnf{}: a sublattice-exchanging rotation makes the detuning odd, its antiunitary combination with time reversal makes the response odd in field, and any easy-axis collinear altermagnet with 
a spectrally isolated doublet inherits the protocol. The magnetic field thus acts as a magnifying lens for magnon spectroscopy: an interaction scale that is quadratic and unsigned at zero field returns linear, 
signed, and resolution-robust, bringing $\mu$eV altermagnetic exchange within reach of present neutron instrumentation.

\begin{acknowledgments}
We thank Jieming Sheng for helpful discussion. This work is supported by the Ministry of Science and Technology of the People's Republic of China (Grant No.~2022YFA1402704), by the National Natural Science Foundation of China (Grants No.~12274187 and No.~12247101), by the Fundamental Research Funds for the Central Universities (Grant No.~lzujbky-2024-jdzx06), and by the Natural Science Foundation of Gansu Province (No.~22JR5RA389).
\end{acknowledgments}

\textit{Data availability.---} The raw data that support the findings of
this article are publicly available on \href{https://github.com/universe9793/MnF2-field-amplified-exchange-data}{GitHub}. 

\bibliography{mnf2-cite}

\clearpage
\setcounter{equation}{0}
\renewcommand{\theequation}{A\arabic{equation}}

\section*{End Matter}

\textit{Appendix A: Hamiltonian, conventions, and first-principles mapping.---}The model used throughout is
\begin{equation}
\mathcal H=\mathcal H_{\rm ex}+\mathcal H_{\rm dip}+\mathcal H_Z,
\label{eq:H}
\end{equation}
\begin{align}
\mathcal H_{\rm ex}&=\frac12\sum_{ij}J_{ij}\,\mathbf S_i\cdot\mathbf S_j,
\label{eq:Hex}\\
\mathcal H_{\rm dip}&=\frac{\mu_0(g\mu_B)^2}{8\pi}\sum_{i\neq j}
\left[\frac{\mathbf S_i\cdot\mathbf S_j}{r_{ij}^3}
-3\frac{(\mathbf S_i\cdot\mathbf r_{ij})(\mathbf S_j\cdot\mathbf r_{ij})}{r_{ij}^5}\right],
\label{eq:Hdip}\\
\mathcal H_Z&=-g\mu_B\,\mathbf B\cdot\sum_i\mathbf S_i,
\label{eq:HZ}
\end{align}
where $\mathbf S_i$ is the spin operator at Mn site $i$, $J_{ij}$ the isotropic exchange between sites $i$ and $j$, $\mathbf r_{ij}=\mathbf r_j-\mathbf r_i$ with $r_{ij}=|\mathbf r_{ij}|$, $\mu_0$ the vacuum permeability, $g$ the Land\'e factor of the Mn moment, $\mu_B$ the Bohr magneton, and $\mathbf B$ the applied field. The sign convention of Eq.~\eqref{eq:Hex} makes a positive $J_{ij}$ antiferromagnetic; the factor $\tfrac12$ compensates the double counting of each bond by the unrestricted sum over $i,j$.

For the mapping, a $20$-Mn and a $28$-Mn magnetic supercell were optimized with \textsc{superhex}~\cite{superhex-PhysRevB.111.144419,superhex-rezaei_benchmarking_2026}, and several hundred collinear magnetic configurations were computed self-consistently in \textsc{vasp}~\cite{dft-vasp-PhysRevB.47.558,dft-vasp-PhysRevB.54.11169,dft-vasp-KRESSE199615} with the projector augmented-wave method~\cite{dft-paw-PhysRevB.50.17953,dft-paw-PhysRevB.59.1758}, the PBE functional~\cite{dft-gga-PhysRevLett.77.3865}, and a Dudarev Hubbard correction on the Mn $d$ shell~\cite{dft-u-PhysRevB.57.1505}. The least-squares inversion resolves $J_1$ through $J_7$ in the $20$-Mn cell and $J_1$ through $J_{12}$ in the $28$-Mn cell.

\textit{Appendix B: Exchange form factor and two-mode projection.---}With $J_{11}(\kvec)$ and $J_{22}(\kvec)$ the momentum-space Fourier components of the exchange within magnetic sublattices $1$ and $2$, the sublattice-odd combination defines the exchange form factor
\begin{equation}
\lam(\kvec)\propto S\bigl[J_{22}(\kvec)-J_{11}(\kvec)\bigr].
\label{eq:formfactor}
\end{equation}
A space-group operation $g$ that interchanges the sublattices obeys $J_{11}(g\kvec)=J_{22}(\kvec)$ and $J_{22}(g\kvec)=J_{11}(\kvec)$, hence $\lam(g\kvec)=-\lam(\kvec)$. For the seventh-neighbor imbalance of \mnf{}, in Cartesian momentum components with $a$, $b$ the in-plane lattice constants,
\begin{equation}
\lam(\kvec)=4S\,\dJ\,\sin(k_xa)\sin(k_yb),
\label{eq:lam_cart}
\end{equation}
equivalent to Eq.~\eqref{eq:lam_rlu} of the main text.

Projected onto the tracked doublet, the quadratic spin-wave problem is the two-mode Hamiltonian
\begin{multline}
\mathcal H_{\rm eff}(\Q,B)=\bar\omega_{\Q}\,I
+\tfrac12\bigl[\Lambda_{\Q}+h_{\Q}(B)\bigr]\sigma_z\\
+d_x(\Q)\sigma_x+d_y(\Q)\sigma_y,
\label{eq:Heff}
\end{multline}
with $\bar\omega_{\Q}=(\omega_++\omega_-)/2$ the doublet center, $I$ the $2\times2$ identity, $\Lambda_{\Q}$ the intrinsic diagonal detuning of Eq.~\eqref{eq:Lambda_conv}, and $\sigma_{x,y,z}$ the Pauli matrices; its eigenvalue splitting is Eq.~\eqref{eq:master} of the main text. The transverse amplitudes are dipolar. In reciprocal space the DDI is the tensor
\begin{equation}
\mathcal D^{\alpha\beta}_{ss'}(\kvec)=\frac{\mu_0(g\mu_B)^2}{4\pi}
{\sum_{\mathbf R}}'
\left[\frac{\delta_{\alpha\beta}}{r^3}-3\frac{r_\alpha r_\beta}{r^5}\right]e^{i\kvec\cdot\mathbf r},
\label{eq:Dtensor}
\end{equation}
with $\mathbf r=\mathbf R+\bm\tau_{s'}-\bm\tau_s$, sublattice indices $s,s'$, Cartesian spin indices $\alpha,\beta$, and the prime excluding the $\mathbf r=0$ self term; the lattice sum is evaluated by Ewald summation. Rotated to the local transverse basis of the ordered state and projected onto the doublet, it yields the complex matrix element $V_{\Q}=d_x(\Q)-id_y(\Q)$ and the projected mixing scale
\begin{equation}
D_{\Q}=2|V_{\Q}|=2\sqrt{d_x^2(\Q)+d_y^2(\Q)}
\label{eq:D_def}
\end{equation}
used in the main text.

\end{document}


\title{Supplemental Material for ``Field-amplified readouts of weak altermagnetic exchange in \mnf''}

\author{Guowen Jiang}
\affiliation{Lanzhou Center for Theoretical Physics, Lanzhou University, Lanzhou 730000, China}
\affiliation{Key Laboratory of Quantum Theory and Applications of MoE, Lanzhou University, Lanzhou 730000, China}
\affiliation{Key Laboratory of Theoretical Physics of Gansu Province and Gansu Provincial Research Center for Basic Disciplines of Quantum Physics, Lanzhou University, Lanzhou 730000, China}

\author{Feilong Wang}
\affiliation{Lanzhou Center for Theoretical Physics, Lanzhou University, Lanzhou 730000, China}
\affiliation{Key Laboratory of Quantum Theory and Applications of MoE, Lanzhou University, Lanzhou 730000, China}
\affiliation{Key Laboratory of Theoretical Physics of Gansu Province and Gansu Provincial Research Center for Basic Disciplines of Quantum Physics, Lanzhou University, Lanzhou 730000, China}

\author{Yunhua Wang}
\affiliation{Lanzhou Center for Theoretical Physics, Lanzhou University, Lanzhou 730000, China}
\affiliation{Key Laboratory of Quantum Theory and Applications of MoE, Lanzhou University, Lanzhou 730000, China}
\affiliation{Key Laboratory of Theoretical Physics of Gansu Province and Gansu Provincial Research Center for Basic Disciplines of Quantum Physics, Lanzhou University, Lanzhou 730000, China}

\author{Fawei Zheng}
\affiliation{Centre for Quantum Physics, Key Laboratory of Advanced Optoelectronic Quantum Architecture and Measurement (MOE), School of Physics, Beijing Institute of Technology, Beijing 100081, China}
\affiliation{Beijing Key Laboratory of Nanophotonics and Ultrafine Optoelectronic Systems, School of Physics, Beijing Institute of Technology, Beijing 100081, China}
\affiliation{International Center for Quantum Materials, Beijing Institute of Technology, Zhuhai 519000, China}

\author{Bin Xi}
\email{binxi@yzu.edu.cn}
\affiliation{College of Physics Science and Technology, Yangzhou University, Yangzhou 225002, China}

\author{Hong-Gang Luo}
\affiliation{Lanzhou Center for Theoretical Physics, Lanzhou University, Lanzhou 730000, China}
\affiliation{Key Laboratory of Quantum Theory and Applications of MoE, Lanzhou University, Lanzhou 730000, China}
\affiliation{Key Laboratory of Theoretical Physics of Gansu Province and Gansu Provincial Research Center for Basic Disciplines of Quantum Physics, Lanzhou University, Lanzhou 730000, China}

\author{Jize Zhao}
\email{zhaojz@lzu.edu.cn}
\affiliation{Lanzhou Center for Theoretical Physics, Lanzhou University, Lanzhou 730000, China}
\affiliation{Key Laboratory of Quantum Theory and Applications of MoE, Lanzhou University, Lanzhou 730000, China}
\affiliation{Key Laboratory of Theoretical Physics of Gansu Province and Gansu Provincial Research Center for Basic Disciplines of Quantum Physics, Lanzhou University, Lanzhou 730000, China}
\maketitle

\section{First-principles total-energy mapping and statistical diagnostics}
\label{sec:s1}

First-principles calculations were performed with the Vienna \textit{ab initio} Simulation Package (VASP)~\cite{dft-vasp-PhysRevB.47.558,dft-vasp-PhysRevB.54.11169,dft-vasp-KRESSE199615}. Electron--ion interactions were described by the projector augmented-wave (PAW) method~\cite{dft-paw-PhysRevB.50.17953,dft-paw-PhysRevB.59.1758}. We used the Perdew--Burke--Ernzerhof (PBE) generalized gradient approximation~\cite{dft-gga-PhysRevLett.77.3865} with a plane-wave cutoff of 520~eV and a Dudarev Hubbard correction on the Mn $d$ orbitals~\cite{dft-u-PhysRevB.57.1505}; the production value is $U_{\mathrm{eff}}=4.8$~eV, with additional scans at $3.8$, $4.3$, $5.3$, and $5.8$~eV. One common structure was relaxed until residual forces were below $0.01$~eV/\AA{} and then fixed for all collinear magnetic configurations. Static energies were computed without spin--orbit coupling on a $\Gamma$-centered $4\times4\times4$ $k$-point mesh with a $10^{-8}$~eV electronic convergence criterion.

For normalized collinear variables $\bar S_i=\pm1$, the energy of configuration $m$ is
\begin{equation}
	E_m=E_0+\frac{1}{2}\sum_{ij}\bar J_{ij}\bar S_i\bar S_j,
\qquad \bar J_{ij}=S^2J_{ij},
\end{equation}
with $S=5/2$. The 20-Mn supercell resolves interactions from $J_1$ through $J_7$, including the split pair $J_{7a,b}$; the 28-Mn supercell extends the mapping through $J_{12}$. In total, 475 collinear configurations enter the overdetermined mapping.

\begin{table}[htbp]
\caption{Diagnostics for the 28-Mn mapping at $U_{\mathrm{eff}}=4.8$~eV. Energy residuals are given per magnetic supercell. RMSE (root-mean-square error of the configuration-energy residuals of the least-squares fit), MAE (mean absolute error), and the maximum absolute residual are in meV.}
\label{tab:fitdiag}
\centering
\small
\begin{tabular*}{\linewidth}{@{\extracolsep{\fill}}cccccc@{}}
\toprule
Configurations & Condition number & RMSE & MAE & Maximum absolute residual & $R^2$ \\
\midrule
235 & 24.68 & 0.02274 & 0.01845 & 0.06290 & 0.99999875 \\
\bottomrule
\end{tabular*}
\end{table}

For each bootstrap replica, the configuration--energy pairs are resampled with replacement to the original sample size, and the full exchange model is refitted. We retain $10^4$ successful replicas for each supercell and each $U_{\mathrm{eff}}$ value. The sign of $\dJ=J_{7b}-J_{7a}$ is positive in every replica at every $U_{\mathrm{eff}}$.

\begin{figure}[htbp]
\centering
\includegraphics[width=0.78\textwidth]{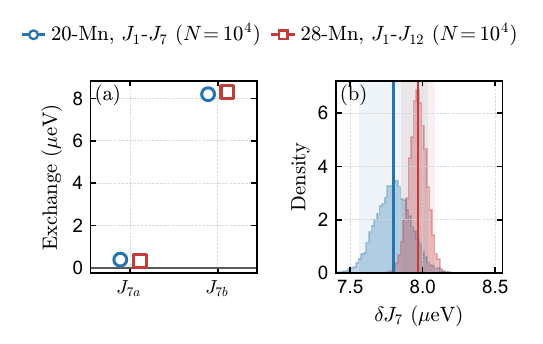}
\caption{(a) Medians and 95\% percentile intervals of $J_{7a}$ and $J_{7b}$ for the 20-Mn mapping from $J_1$ through $J_7$ and the 28-Mn mapping from $J_1$ through $J_{12}$. (b) Normalized distributions of $\dJ=J_{7b}-J_{7a}$. Panel (b) uses the Letter notation $\delta J_7$.}
\label{fig:bootdetail}
\end{figure}

\section{Exchange parameters, $U_{\mathrm{eff}}$ stability, and the $J_1$ anomaly}
\label{sec:s2}

Table~\ref{tab:exchange} lists the production exchange parameters; Table~\ref{tab:uscan} gives the five-point $U_{\mathrm{eff}}$ scan. The imbalance $\dJ$ decreases smoothly from $11.4$ to $5.4~\mu$eV across the physical $U_{\mathrm{eff}}$ range but never changes sign; the path hierarchy $J_{7b}\gg J_{7a}$ ($\sim$22--27:1) is likewise stable.
In terms of the field-linear readouts of the main text, this systematic translates into a compensation-field window $|B^{*}(\mathbf{Q}_A)| \simeq
0.23$--$0.49$~T across the physical $U_{\mathrm{eff}}$ range, always well within reach of standard cryomagnets, while the sign of $B^{*}$, fixed by
the sign of $\delta J_7$, is $U_{\mathrm{eff}}$-independent.

\begin{table}[htbp]
\caption{First-principles 28-Mn production parameters at $U_{\mathrm{eff}}=4.8$~eV. Values are in $\mu$eV.}
\label{tab:exchange}
\centering
\begin{tabular}{lrrr}
\toprule
Parameter & Least-squares fit & Bootstrap median & 95\% interval \\
\midrule
$J_1$   & 28.598  & 28.595  & [28.434, 28.756] \\
$J_2$   & 338.879 & 338.878 & [338.763, 338.999] \\
$J_3$   & 2.865   & 2.863   & [2.775, 2.958] \\
$J_4$   & 5.461   & 5.459   & [5.394, 5.528] \\
$J_5$   & $-0.512$ & $-0.512$ & [$-0.559$, $-0.466$] \\
$J_6$   & 4.823   & 4.824   & [4.678, 4.987] \\
$J_{7a}$& 0.336   & 0.331   & [0.210, 0.447] \\
$J_{7b}$& 8.301   & 8.298   & [8.170, 8.423] \\
$J_8$   & 0.283   & 0.282   & [0.188, 0.374] \\
$J_9$   & $-0.045$ & $-0.045$ & [$-0.109$, 0.018] \\
$J_{10}$& $-0.154$ & $-0.151$ & [$-0.289$, $-0.011$] \\
$J_{11}$& 0.122   & 0.121   & [$-0.025$, 0.264] \\
$J_{12}$& 0.032   & 0.032   & [$-0.014$, 0.077] \\
\bottomrule
\end{tabular}
\end{table}

\begin{table}[htbp]
\caption{28-Mn exchange parameters from the five-point $U_{\mathrm{eff}}$ scan. Values are in $\mu$eV.}
\label{tab:uscan}
\centering
\small
\begin{tabular*}{\linewidth}{@{\extracolsep{\fill}}lrrrrr@{}}
\toprule
Parameter & \multicolumn{5}{c}{$U_{\mathrm{eff}}$ (eV)} \\
\cmidrule(lr){2-6}
& 3.8 & 4.3 & 4.8 & 5.3 & 5.8 \\
\midrule
$J_1$    & 68.678  & 46.866  & 28.598  & 13.305  & 0.529 \\
$J_2$    & 438.906 & 385.895 & 338.879 & 296.987 & 259.503 \\
$J_3$    & 3.584   & 3.196   & 2.865   & 2.579   & 2.329 \\
$J_4$    & 7.013   & 6.183   & 5.461   & 4.829   & 4.272 \\
$J_5$    & $-0.607$ & $-0.562$ & $-0.512$ & $-0.460$ & $-0.409$ \\
$J_6$    & 6.003   & 5.377   & 4.823   & 4.329   & 3.887 \\
$J_{7a}$ & 0.445   & 0.385   & 0.336   & 0.295   & 0.260 \\
$J_{7b}$ & 11.805  & 9.925   & 8.301   & 6.890   & 5.659 \\
$J_8$    & 0.489   & 0.372   & 0.283   & 0.216   & 0.164 \\
$J_9$    & $-0.062$ & $-0.053$ & $-0.045$ & $-0.039$ & $-0.033$ \\
$J_{10}$ & $-0.219$ & $-0.183$ & $-0.154$ & $-0.129$ & $-0.109$ \\
$J_{11}$ & 0.175   & 0.146   & 0.122   & 0.103   & 0.087 \\
$J_{12}$ & 0.053   & 0.041   & 0.032   & 0.025   & 0.019 \\
\midrule
$\dJ$ & 11.360 & 9.540 & 7.965 & 6.595 & 5.399 \\
\bottomrule
\end{tabular*}
\end{table}

\emph{The $J_1$ sign anomaly.} With the convention $\mathcal H=+\tfrac12\sum_{ij}J_{ij}\mathbf S_i\!\cdot\!\mathbf S_j$, positive $J_1$ in Table~\ref{tab:uscan} is antiferromagnetic, whereas neutron-scattering parameterizations usually assign a weakly ferromagnetic sign to $J_1$~\cite{mnf-neutron-doi:10.1139/P10-081,mnf-neutron-osti_4092064,pins-faure2025altermagnetismrevealedpolarizedneutrons,magnon-absencePhysRevLett.134.226702}. One explanation is a near cancellation between antiferromagnetic superexchange and ferromagnetic direct exchange: when the net coupling is small, a slight functional- or $U_{\mathrm{eff}}$-dependent error in their relative weights can change its sign. In the present scan $J_1$ decreases monotonically from $68.7$ to $0.5~\mu$eV, described by
\begin{equation}
J_1(U_{\mathrm{eff}})=\frac{C}{U_{\mathrm{eff}}}-J_0,
\qquad
C=751.8~\mu\mathrm{eV\,eV},\quad
J_0=128.6~\mu\mathrm{eV},
\end{equation}
with $R^2=0.9996$ and RMSE $=0.51~\mu$eV---consistent with a near cancellation of contributions with different $U_{\mathrm{eff}}$ dependences. The same anomaly appears in the independent linear-response calculation of Ref.~\cite{mnf-band-solovyev2026altermagnetismmnf2bandsplitting}, whose $c$-axis intra-sublattice coupling is also antiferromagnetic against the experimental ferromagnetic assignment: the discrepancy is a functional-level systematic of the weak, nearly canceling $J_1$, not a defect of the mapping. By contrast, $\dJ$ stays positive over the full range. A controlled spin-wave check (replacing $J_1$ by two published ferromagnetic values and by zero, all else fixed) changes branch separations by less than $10^{-6}~\mu$eV and mode chiralities by at most $1.1\times10^{-5}$: the $J_1$ sign does not affect any conclusion of this work.

\emph{Symmetry-complete extension.} Splitting the eighth-neighbor shell into $J_{8a}$ and $J_{8b}$ gives $\delta J_8=J_{8b}-J_{8a}=0.0315~\mu$eV with $\mathrm{CI}_{95\%}=[-0.098,0.166]~\mu$eV and $|\delta J_8/\dJ|=0.004$: the orbit is symmetry-allowed but unresolved, and we use $J_{8a}=J_{8b}\equiv J_8$ in the spin-wave calculations.

\section{Spin-wave theory, Ewald dipolar summation, and zero-parameter anchors}
\label{sec:s3}

The mapped exchange model is propagated through linear spin-wave theory with the long-range dipole--dipole interaction (DDI) evaluated by an Ewald procedure~\cite{ewald-PhysRevB.70.174426}. No phenomenological single-ion anisotropy is introduced: the Ewald-summed DDI alone opens the zone-center gap, and this choice keeps the separation between exchange imbalance and dipolar dressing explicit rather than absorbing dipolar effects into an effective local anisotropy.

Two zero-parameter anchors test the forward machinery against published experiments, with no quantity adjusted:
\begin{enumerate}
\item \emph{Zone-center gap.} The calculated DDI gap is $1.145$~meV, against the experimental scale of about $1.08$~meV~\cite{mnf-neutron-doi:10.1139/P10-081,mnf-neutron-osti_4092064}.
\item \emph{Splitting at the identical momentum.} At $\Q=(\tfrac12,\tfrac12,1)$---exactly the momentum at which experiment reports the maximal splitting $178\pm3~\mu$eV~\cite{pins-faure2025altermagnetismrevealedpolarizedneutrons}---the finite-momentum Ewald calculation gives
\begin{equation}
\Omega_{\rm Ewald}=179.419~\mu\mathrm{eV}.
\end{equation}
An independent real-space spherical-cutoff summation converges to $179.364~\mu$eV at cutoff radius $60$~\AA, a $0.055~\mu$eV cross-check difference, and the Ewald result is stable as the separation parameter varies over $\eta=0.10$--$0.20$~\AA$^{-1}$. The calculated central value differs from the experimental central value by $1.4~\mu$eV, i.e.\ $0.47$ experimental standard deviations.
\end{enumerate}
This same-momentum agreement validates the dipolar lattice summation and the forward spin-wave calculation; it does not by itself determine the sign of $\dJ$, which is the subject of the field protocol.

\begin{figure}[htbp]
\centering
\includegraphics[width=0.88\textwidth]{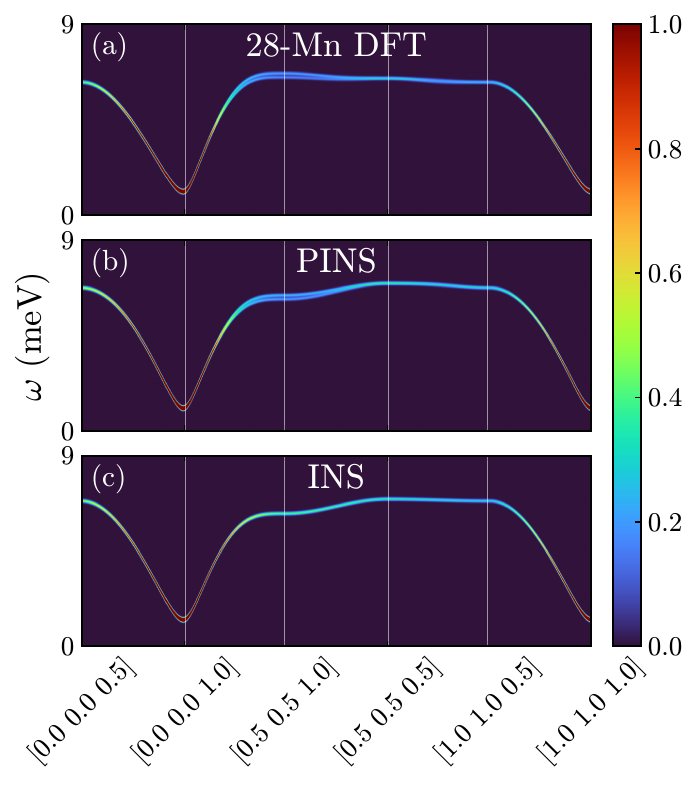}
\caption{Unpolarized dynamic structure factor $I_0(\Q,\omega)$ along a common momentum path. (a) Present first-principles model. (b,c) Two published phenomenological parameterizations~\cite{pins-faure2025altermagnetismrevealedpolarizedneutrons,magnon-absencePhysRevLett.134.226702}. All spectra are calculated at $T=0$, use $\Gamma_{\mathrm{FWHM}}=0.12$~meV, and are normalized independently.}
\label{fig:unpolarized_compare}
\end{figure}

Figure~\ref{fig:unpolarized_compare} compares the calculated unpolarized structure factor with two published phenomenological descriptions along a common momentum path.

\subsection{Spin--orbit single-ion anisotropy from first principles}
\label{sec:s3-sia}

\emph{Method and frames.} The single-ion anisotropy (SIA) of Mn$^{2+}$ is computed by the four-state total-energy mapping in VASP on a $2\times2\times3$ supercell, with the spin--orbit step evaluated non-self-consistently from initial spin densities prepared by Spinss~\cite{dft-spinss-YANG2026110297}. For each Mn site $i$ the mapping resolves the full quadratic anisotropy tensor $\mathbf A^{(i)}$ of the site term $\sum_{\alpha\beta}A^{(i)}_{\alpha\beta}S_{\alpha}S_{\beta}$, in a site-local Cartesian frame whose axes coincide with the local $mmm$ principal axes of the Mn site. For an $mmm$ site an in-plane anisotropy $A_{xx}\neq A_{yy}$ is symmetry-allowed, while the off-diagonal elements vanish for the ideal symmetry; the computed off-diagonal elements, at most $0.22~\mu$eV on either sublattice, are therefore a direct residual of the present numerical precision.

\emph{Site tensors.} In units of $\mu$eV,
\begin{equation}
\mathbf A^{(1)}=
\begin{pmatrix}
+4.15 & +0.22 & +0.03\\
+0.22 & -1.43 & +0.01\\
+0.03 & +0.01 & -2.73
\end{pmatrix},
\qquad
\mathbf A^{(2)}=
\begin{pmatrix}
+2.49 & -0.21 & -0.01\\
-0.21 & -4.21 & -0.01\\
-0.01 & -0.01 & +1.72
\end{pmatrix}.
\label{eq:s3-sia-tensors}
\end{equation}
With the axial and in-plane combinations $D_i=A^{(i)}_{zz}-\tfrac12\bigl[A^{(i)}_{xx}+A^{(i)}_{yy}\bigr]$ and $E_i=\tfrac12\bigl[A^{(i)}_{xx}-A^{(i)}_{yy}\bigr]$, these give
\begin{equation}
D_1=-4.09~\mu\mathrm{eV},\quad E_1=+2.79~\mu\mathrm{eV};
\qquad
D_2=+2.58~\mu\mathrm{eV},\quad E_2=+3.35~\mu\mathrm{eV}.
\end{equation}
Neither single-site value is, by itself, the physical SIA: the two Mn sites are mapped onto each other by the sublattice-exchanging screw, so the physical statement is the comparison of the two sites in a common frame.

\emph{$C_4$ mapping and symmetrization.} Rotating site 2 into the frame of site 1 with the fourfold rotation transforms the tensor elements as
\begin{equation}
A_{xx}\leftrightarrow A_{yy},\qquad
A_{xy}\to-A_{xy},\qquad
A_{xz}\to-A_{yz},\qquad
A_{yz}\to-A_{xz},
\label{eq:s3-sia-map}
\end{equation}
giving
\begin{equation}
\mathbf A^{(2\to1)}=
\begin{pmatrix}
-4.21 & +0.21 & +0.01\\
+0.21 & +2.49 & +0.01\\
+0.01 & +0.01 & +1.72
\end{pmatrix}~\mu\mathrm{eV},
\qquad
D_2=+2.58~\mu\mathrm{eV},\quad E_2=-3.35~\mu\mathrm{eV}:
\end{equation}
the axial parameter is invariant under the mapping while the in-plane parameter reverses sign. In an exact calculation $\mathbf A^{(1)}$ and $\mathbf A^{(2\to1)}$ would coincide; their residual difference measures the numerical uncertainty of the mapping. The symmetrized tensor
\begin{equation}
\mathbf A_{\rm sym}=\tfrac12\bigl[\mathbf A^{(1)}+\mathbf A^{(2\to1)}\bigr]=
\begin{pmatrix}
-0.03 & +0.22 & +0.02\\
+0.22 & +0.53 & +0.01\\
+0.02 & +0.01 & -0.50
\end{pmatrix}~\mu\mathrm{eV}
\end{equation}
yields
\begin{equation}
D_{\rm sym}=-0.75\pm3.34~\mu\mathrm{eV},
\qquad
E_{\rm sym}=-0.28\pm3.07~\mu\mathrm{eV},
\end{equation}
the quoted uncertainty being half the spread between the two symmetry-equivalent sites. The electronic SIA is thus compatible with zero at the present first-principles precision, with the conservative bounds
\begin{equation}
|D_{\rm elec}|\lesssim4.1~\mu\mathrm{eV},
\qquad
|E_{\rm elec}|\lesssim3.4~\mu\mathrm{eV}
\label{eq:s3-sia-bounds}
\end{equation}
set by the largest single-site magnitudes; the smallness of all off-diagonal elements ($\le0.22~\mu$eV, well below the diagonal anisotropy scale) is an independent internal consistency check.

\begin{figure}[htbp]
\centering
\includegraphics[width=0.95\textwidth]{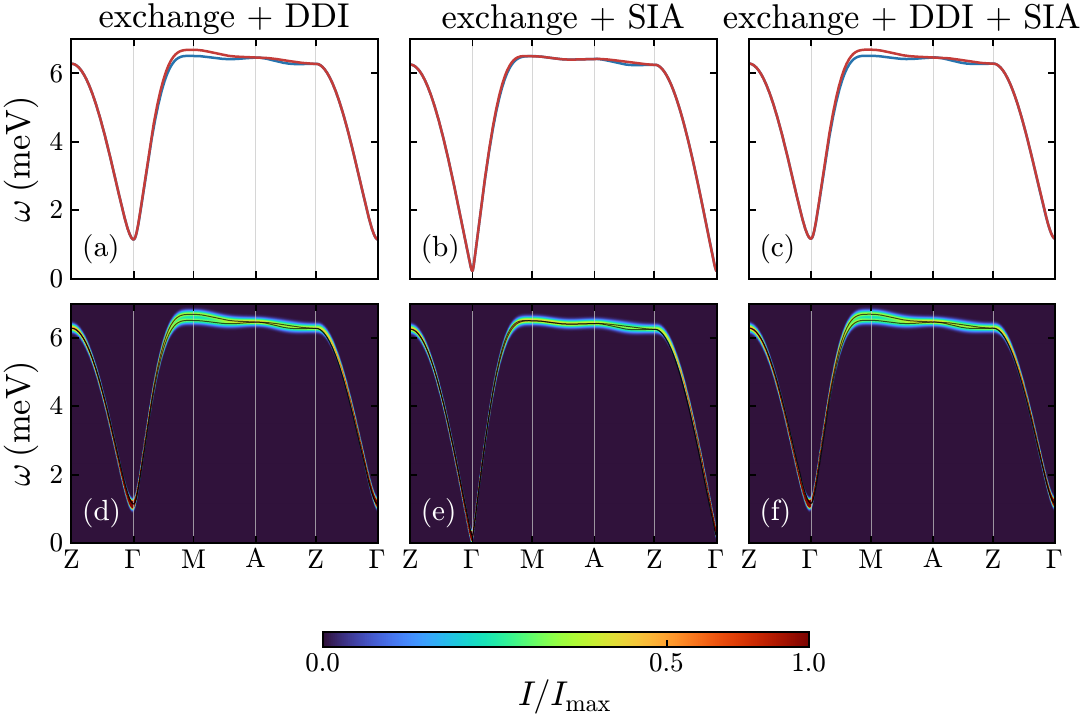}
\caption{Spin-wave dispersions (top row) and resolution-convolved unpolarized INS intensity $I/I_{\max}$ (bottom row) along $Z$--$\Gamma$--$M$--$A$--$Z$--$\Gamma$ for three models: exchange with DDI [(a),(d), the working model of this work], exchange with the first-principles single-ion anisotropy (SIA) [(b),(e)], and exchange with both [(c),(f)]. Colors in the top row distinguish the two magnon branches. The SIA alone opens a zone-center gap well below the observed $\sim1$~meV scale; together with the DDI the gap is $\simeq1.17$~meV, and the field readouts of the Letter are essentially unchanged.}
\label{fig:sia-compare}
\end{figure}

\emph{Consequences for the spectrum.} Figure~\ref{fig:sia-compare} propagates a single-ion term at the few-$\mu$eV scale of these bounds through the same spin-wave machinery: alone it opens a zone-center gap well below the observed $\sim1$~meV scale, and combined with the Ewald-summed DDI it shifts the calculated gap from $1.15$ to approximately $1.17$~meV. Because the site tensors of the two sublattices map onto each other under the sublattice-exchanging screw, the single-ion term is sublattice-even: it contributes nothing to the sublattice-odd magnon detuning, its inclusion leaves the compensation fields and the odd-in-field contrast of Secs.~\ref{sec:s5} and~\ref{sec:s7} essentially unchanged, and the DDI-only model is retained throughout.

\section{Zero-field decomposition and the dipolar-mixing-free window}
\label{sec:s4}

\emph{Quadrature composition.} At a generic momentum the exchange imbalance and the dipolar mixing combine in quadrature. At $\Q=(0.25,0.25,1)$ the exchange-only splitting is $\Omega_{\rm ex}=79.6~\mu$eV, the symmetry-restored (equal-$J_7$) model with DDI gives $\Omega_{\rm DDI}=121.6~\mu$eV, and
\begin{equation}
\Omega_{\rm full}=\sqrt{\Omega_{\rm ex}^2+\Omega_{\rm DDI}^2}=145.3~\mu\mathrm{eV},
\end{equation}
agrees with the full numerical result at the stated precision. In the two-mode language of Sec.~\ref{sec:s5} these three quantities are $|\Lambda|$, $D$, and $\Omega(0)$; modulo reciprocal-lattice vectors, $(0.25,0.25,1)$ is the $C_{4z}$ partner of the Letter's $\Q_A=(0.25,0.75,0)$, and all three are partner-even, so the same values hold at the pair $\Q_A$, $\Q_{A'}$ used in the Letter. The same ratio gives a diluted mode chirality $|\chi|\simeq0.55$, against $0.54$--$0.56$ from the Bogoliubov eigenvectors: DDI enhances the visible separation while diluting chirality. Because the composition is quadratic, the zero-field splitting is an even function of $\dJ$---the structural obstruction that motivates the compensation-field protocol of Sec.~\ref{sec:s5}.

\emph{Dipolar-mixing-free window.} Along the momentum line $(H,H,\tfrac12)$ the projected transverse dipolar mixing can vanish by symmetry within the adopted Hamiltonian, while the full DDI remains, leaving an exchange-driven splitting, e.g.\ $\approx80~\mu$eV at $(0.75,0.75,0.5)$ with mode chirality $|\chi|\approx0.86$. A zero-field measurement in this window is a complementary check on $|\dJ|$---though, like every zero-field observable, it remains even in $\dJ$ and carries no sign information.

\section{Conventions, mode ordering, and the compensation-field derivation}
\label{sec:s5}

This section fixes, exactly once, the bond, momentum, domain, and mode-ordering conventions used in the Letter, derives the compensation field $B^\ast(\Q)$ and the odd combination $\mathcal{C}(B)$ at the level of the full Bogoliubov--de~Gennes (BdG) spin-wave problem, and writes the chain from the microscopic sign of $\dJ$ to the polarity of $B^\ast_A=-B^\ast_{A'}$. Every sign statement in the Letter inherits these conventions.

\subsection{Conventions and the observable}

\emph{Lattice, domains, field.} We use the right-handed crystal axes $(a,b,c)$ of the rutile cell. Magnetic sublattice $1$ sits at $(0,0,0)$ with reference magnetic moment along $+\hat c$; 
sublattice $2$ sits at $(\tfrac12,\tfrac12,\tfrac12)$ with reference magnetic moment along $-\hat c$. This magnetic configuration is the domain $D_+$; reversing every moment gives the time-reversed N\'eel domain $D_-$. The field is applied along the easy axis, $\mathbf B=B\hat c$, with $B>0$ meaning $+\hat c$.

\emph{Exchange labels.} Throughout the paper,
\begin{equation}
\dJ\equiv J_{7b}-J_{7a},
\label{eq:s5-dj7}
\end{equation}
where, on sublattice $1$, $J_{7a}$ couples along the in-plane bond vector $(1,-1,0)$ and $J_{7b}$ along $(1,1,0)$---the path bridged by two fluorine ions. On sublattice $2$ the two diagonal directions are interchanged. The $J_{7a}/J_{7b}$ labels of Ref.~\cite{pins-faure2025altermagnetismrevealedpolarizedneutrons} are geometrically opposite to this definition, and so are the labels of the \fef{} literature model~\cite{pins-g6dt-rf8c}: literature parameters must be relabeled ($a\leftrightarrow b$) before any sign comparison; the literal table order carries no meaning across conventions (Secs.~\ref{sec:s8} and \ref{sec:s9}).

\emph{Momenta and observable.} The partner momenta are
\begin{equation}
\Q_A=(0.25,0.75,0),\qquad
\Q_{A'}=(-0.75,0.25,0)=C_{4z}\Q_A.
\end{equation}
At each, $\Omega_q(B)=\omega_{q,+}(B)-\omega_{q,-}(B)$ is the energy difference of the target doublet. The primary spectroscopic observable of the Letter is the compensation field $B^\ast_q=\argmin_B\Omega_q(B)$; the odd combination used as a cross-check is
\begin{equation}
\mathcal{C}(B)=\Omega_A^2(B)-\Omega_{A'}^2(B).
\label{eq:s5-crit}
\end{equation}
Under the conventions above, the adopted \mnf{} model obeys
\begin{equation}
\dJ>0\;\Longrightarrow\;B_A^*<0,\quad B_{A'}^*>0,\quad \frac{d\mathcal{C}}{dB}>0,
\label{eq:s5-polarity}
\end{equation}
with $B_q^*$ the field at which $\Omega_q$ is minimal. Exchanging $\Q_A\leftrightarrow\Q_{A'}$, reversing the field axis, or reversing the domain each flips this polarity relation; a figure caption is therefore complete only if it states the domain, the momentum labels, and the field convention.

\subsection{Two-mode effective model and partner identities}

With the domain fixed, the target doublet at momentum $q$ is described by
\begin{equation}
H_q(B)=\omega_{0q}(B)\,\openone
+\frac{\Lambda_q+2g_q\mu_BB}{2}\,\sigma_z
+d_{xq}\sigma_x+d_{yq}\sigma_y ,
\label{eq:s5-2mode}
\end{equation}
where $\Lambda_q$ is the signed diagonal detuning in the fixed two-mode basis, $2g_q\mu_BB$ is the relative Zeeman detuning, and $d_{xq},d_{yq}$ collect the mode mixing---in the adopted rutile models, supplied by the dipolar interaction. With $D_q^2=4(d_{xq}^2+d_{yq}^2)$,
\begin{equation}
\Omega_q^2(B)=\bigl(\Lambda_q+2g_q\mu_BB\bigr)^2+D_q^2 :
\label{eq:s5-hyperbola}
\end{equation}
the mixing sets the floor of the avoided crossing; the signed detuning sets the position and field polarity of the valley.

For the $C_{4z}$-partner pair, symmetry gives
\begin{equation}
\Lambda_{A'}=-\Lambda_A,\qquad g_{A'}=g_A\equiv g,\qquad D_{A'}=D_A\equiv D ,
\label{eq:s5-partner}
\end{equation}
hence the mirror relation and the dip fields
\begin{equation}
\Omega_A(B)=\Omega_{A'}(-B),\quad
B_A^*=-\frac{\Lambda_A}{2g\mu_B},\quad
B_{A'}^*=+\frac{\Lambda_A}{2g\mu_B}.
\label{eq:s5-mirror}
\end{equation}

\subsection{The mode-ordering sign and the complete polarity chain}

The pure-exchange dispersion contains the altermagnetic term
\begin{equation}
\lam(\kvec)=4S\,\dJ\sin(k_xa)\sin(k_yb).
\label{eq:s5-lam}
\end{equation}
Here $\lam$ is the signed exchange term of the dispersion formula, while $\Lambda_q$ of Eq.~(\ref{eq:s5-2mode}) is the diagonal detuning \emph{in the fixed ordering of the two-mode basis}. These are not the same signed quantity. The adopted BdG basis and branch ordering give
\begin{equation}
\boxed{\;\Lambda_q=-\lam(\Q_q)\;}
\label{eq:s5-ordering}
\end{equation}
The minus sign originates in the ordering of the two diagonal basis vectors, not in an arbitrary Bloch phase: exchanging the two basis vectors flips $\sigma_z$, and with it $\Lambda_q$ \emph{and} the Zeeman detuning simultaneously, leaving the physical valley positions unchanged. The ordering-independent physical invariant is the product $\Lambda_q\,\partial_B h_q$---equivalently, the slope of $\mathcal{C}(B)$.

Evaluating Eq.~(\ref{eq:s5-lam}) at the partner momenta:
\begin{align}
&\Q_A:\;\sin(2\pi\cdot0.25)\sin(2\pi\cdot0.75)=-1
\;\Rightarrow\;\lam(\Q_A)=-4S\dJ,\quad\Lambda_A=+4S\dJ;\nonumber\\
&\Q_{A'}:\;\sin(-2\pi\cdot0.75)\sin(2\pi\cdot0.25)=+1
\;\Rightarrow\;\lam(\Q_{A'})=+4S\dJ,\quad\Lambda_{A'}=-\Lambda_A .
\end{align}
The complete sign chain, in the conventions of Sec.~S5.1, is therefore
\begin{equation}
\boxed{\;
\dJ>0
\;\Rightarrow\;
\lam(\Q_A)<0
\;\Rightarrow\;
\Lambda_A>0
\;\Rightarrow\;
B_A^*<0
\;\Rightarrow\;
\frac{d\mathcal{C}}{dB}>0
\;}
\label{eq:s5-chain}
\end{equation}
A reader who substitutes $\Q_A$ into Eq.~(\ref{eq:s5-lam}) and identifies the result directly with $\Lambda_A$ obtains the opposite polarity; Eq.~(\ref{eq:s5-ordering}) is the step that must not be skipped.

\subsection{Exact dipolar cancellation and the slope}

Subtracting the squared splittings of Eq.~(\ref{eq:s5-hyperbola}) for the partner pair, the common mixing term $D^2$ cancels exactly:
\begin{equation}
\mathcal{C}(B)=\Omega_A^2(B)-\Omega_{A'}^2(B)=8g\mu_B\Lambda_A B,
\qquad
\frac{d\mathcal{C}}{dB}=8g\mu_B\Lambda_A=32\,g\mu_B S\,\dJ .
\label{eq:s5-slope}
\end{equation}
The sign of the slope reads the sign of $\dJ$ relative to the calibrated majority domain (Sec.~S5.6); its magnitude reads the scale of the imbalance.

The strict linearity of Eq.~(\ref{eq:s5-slope}) is a statement of the reduced two-mode model. The full linear-spin-wave/BdG calculation guarantees the mirror relation and the odd parity $\mathcal{C}(-B)=-\mathcal{C}(B)$ exactly (Sec.~S5.5); mixing with modes outside the target doublet can add higher odd orders $B^3,B^5,\ldots$, but no symmetry-forbidden even term. For the production \mnf{} parameters the full BdG result is linear to within numerical precision over $|B|\le1$~T, with
\begin{equation}
\frac{d\mathcal{C}}{dB}=0.074~\mathrm{meV^2/T},
\end{equation}
matching the two-mode expression. In the equal-$J_7$ reference, $\Lambda_A=0$: both valleys sit at zero field and $\mathcal{C}(B)\equiv0$ identically---while each individual splitting still grows with field.

\subsection{Derivation at the full BdG level and applicability conditions}

The statements above follow from symmetry alone, at the level of the full spin-wave problem. Consider a collinear, compensated magnet with stable N\'eel axis $\hat{\mathbf n}$,
\begin{equation}
\mathcal H=\mathcal H_{\rm ex}+\mathcal H_{\rm ani}+\mathcal H_{\rm dip}+\mathcal H_Z ,
\end{equation}
with isotropic exchange $\mathcal H_{\rm ex}=\tfrac12\sum_{ij}J_{ij}\mathbf S_i\!\cdot\!\mathbf S_j$, any single-ion or exchange anisotropy $\mathcal H_{\rm ani}$ preserving the magnetic symmetry, the point-dipole interaction $\mathcal H_{\rm dip}$, and $\mathcal H_Z=-g\mu_B\mathbf B\cdot\sum_i\mathbf S_i$ with $\mathbf B\parallel\hat{\mathbf n}$. The derivation assumes a fixed lattice and fixed collinear structure; magnetoelastic renormalization and the spin-flopped phase are outside its scope.

\emph{Odd transformation of the detuning.} Let $g=\{R_g|\bm\tau_g\}$ be a space-group operation interchanging the two magnetic sublattices (for the rutiles, the screw rotation; the fractional translation enters only Bloch phases). The intra-sublattice exchange Fourier components obey $J_{11}(g\kvec)=J_{22}(\kvec)$ and vice versa, so any detuning built from their difference,
\begin{gather}
\lam(\kvec)=S\bigl[J_{22}(\kvec)-J_{11}(\kvec)\bigr],\nonumber\\
\boxed{\;\lam(g\kvec)=-\lam(\kvec)\;}
\label{eq:s5-odd}
\end{gather}
is odd between partner momenta. [For the \mnf{} $J_7$ pair this reduces to Eq.~(\ref{eq:s5-lam}).] The construction requires no $d$-wave or $g$-wave labels---only that the chosen operation reverses the sign of the target detuning. If several microscopic interactions share the same odd symmetry, the experiment measures their summed projection onto the target doublet, not any single one of them (see below).

\emph{Antiunitary mirror relation.} The operation $g$ interchanges the antiparallel sublattices and therefore maps a given N\'eel domain to its reverse; the domain-preserving operation is the antiunitary combination $\mathcal A=\Theta g$, with $\Theta$ time reversal. Since $R_g$ preserves the field axis while $\Theta$ reverses $B$, the stable bosonic BdG dynamical matrix $\mathcal M(\kvec,B)=\Sigma_3\mathcal H_{\rm BdG}(\kvec,B)$ obeys
\begin{equation}
\mathcal M(g\kvec,B)=\mathcal U_g\,\mathcal M(\kvec,-B)^{*}\,\mathcal U_g^{-1},
\end{equation}
with $\mathcal U_g$ paraunitary, so the positive-frequency spectra coincide: $\{\omega_n(g\kvec,B)\}=\{\omega_n(\kvec,-B)\}$. If the target doublet is separated from all other branches and can be tracked continuously in field, its splitting obeys the exact mirror relation
\begin{equation}
\boxed{\;\Omega_{A'}(B)=\Omega_A(-B)\;}
\qquad\Longrightarrow\qquad
\mathcal{C}(-B)=-\mathcal{C}(B),
\label{eq:s5-bdgmirror}
\end{equation}
a full-BdG symmetry statement, independent of the two-mode reduction.

\emph{Two-mode projection.} For two isolated, stable modes of positive paraunitary norm, projection onto their positive-energy subspace---using the bosonic BdG metric, not the Euclidean $2\times2$ sub-block---yields Eq.~(\ref{eq:s5-2mode}) with a general Zeeman detuning $h_q(B)$ and field-dependent mixings $D_{xq}(B),D_{yq}(B)$, and
\begin{equation}
\Omega_q^2(B)=\bigl[\Lambda_q+h_q(B)\bigr]^2+D_q^2(B).
\end{equation}
The squared---not linear---appearance of the mixing is why the dipolar term cannot be subtracted from the visible splitting as a fixed background, and why the cancellation must be engineered in the $\Omega^2$ variable. In the adopted rutile models $h(B)=2g\mu_BB$ holds linearly in the scanned range; for general mixed modes $g_{\rm eff}$ is the difference of the diabatic mode moments and need not be constant.

\emph{Scope.} The dipolar interaction is a tensor; ``even'' is not a property of DDI in general. In the adopted models it contributes only the partner-even transverse mixing at the target momenta, which is what cancels in $\mathcal{C}(B)$. Most generally, the criterion measures the total sublattice-odd diagonal detuning of the target doublet: any interaction sharing the odd symmetry of Eq.~(\ref{eq:s5-odd})---including a possible odd diagonal dipolar part in another material---would enter $\Lambda_q$ additively. Attributing the measured $\Lambda_A$ uniquely to $J_{7b}-J_{7a}$ therefore invokes the adopted isotropic-exchange-plus-dipolar Hamiltonian; it does not require reintroducing the microscopic $J$'s into the fit.

\emph{Applicability conditions.} Collecting the assumptions, the criterion applies to a target doublet when: (1) a space operation $g$ interchanges the sublattices with $\Lambda_{g\kvec}=-\Lambda_{\kvec}$; (2) the antiunitary $\Theta g$ preserves the chosen domain and maps $(\kvec,B)\to(g\kvec,-B)$; (3) the partner momenta share the effective Zeeman coefficient; (4) the transverse mixing is partner-even, $D_{g\kvec}^2=D_{\kvec}^2$; (5) the doublet is spectrally isolated and continuously trackable in field; (6) no structural transition, magnetoelastic renormalization, or additional symmetry-odd term intervenes in the scanned range. Conditions (1)--(3) give the exact mirror relation and odd parity; (4)--(5) are needed for the quantitative cancellation of the common mixing in $\mathcal{C}(B)$.

\subsection{Domain mixtures and the majority-domain bit}

For a sample with volume fraction $p$ of domain $D_+$, the incoherent two-domain spectrum obeys the exact degeneracy
\begin{equation}
S(p,\dJ)=S(1-p,-\dJ),
\label{eq:s5-z2}
\end{equation}
because a domain flip acts on the criterion exactly as a sign flip of the exchange imbalance [$\Lambda_q(D_-)=-\Lambda_q(D_+)$, hence $\mathcal{C}_{D_-}(B)=-\mathcal{C}_{D_+}(B)$]. At the level of the likelihood, $(p,\Lambda_A)\leftrightarrow(1-p,-\Lambda_A)$ are exactly equivalent optima; when both float, the fit must carry the majority-domain constraint $p>\tfrac12$, with the calibrated majority domain defined as $D_+$. The scan then determines the sign of $\dJ$ \emph{relative to the majority N\'eel domain}---any $p>0.5$ returns the same sign---and converting to the absolute crystallographic sign costs a single binary bit: which domain is the majority. Standard piezomagnetic domain selection is known in \mnf{}~\cite{borovikromanov1960piezo}; independently, Ref.~\cite{pins-faure2025altermagnetismrevealedpolarizedneutrons} demonstrated an $85\pm5\%$ majority-domain population after cooling through $T_N$. The precise value of $p$ is not needed for the sign; it directly scales the amplitude of the odd signal and must be propagated (or fitted jointly) when $|\dJ|$ is extracted. For field-cooled or stress-trained samples, the experimental record must state the training direction and the N\'eel vector of the majority domain; the raw result is then ``sign of $\dJ$ relative to the calibrated majority domain,'' converted to $J_{7b}-J_{7a}$ through the crystallographic path definitions of Sec.~S5.1. Correspondingly $B^\ast(\Q)$ in the Letter is defined relative to the majority domain.

\subsection{Additional partner-momentum pairs}
\label{sec:s5-family}

The choice of the working pair $\Q, \Q'(=C_{4z}\Q)$ is not unique. To map out the momentum space accessible in finite-field measurements, we calculate the full BdG spectra for two sets of $C_{4z}$-related partner momenta in the $l=0$ plane.

\begin{figure}[htbp]
\centering
\includegraphics[width=0.95\textwidth]{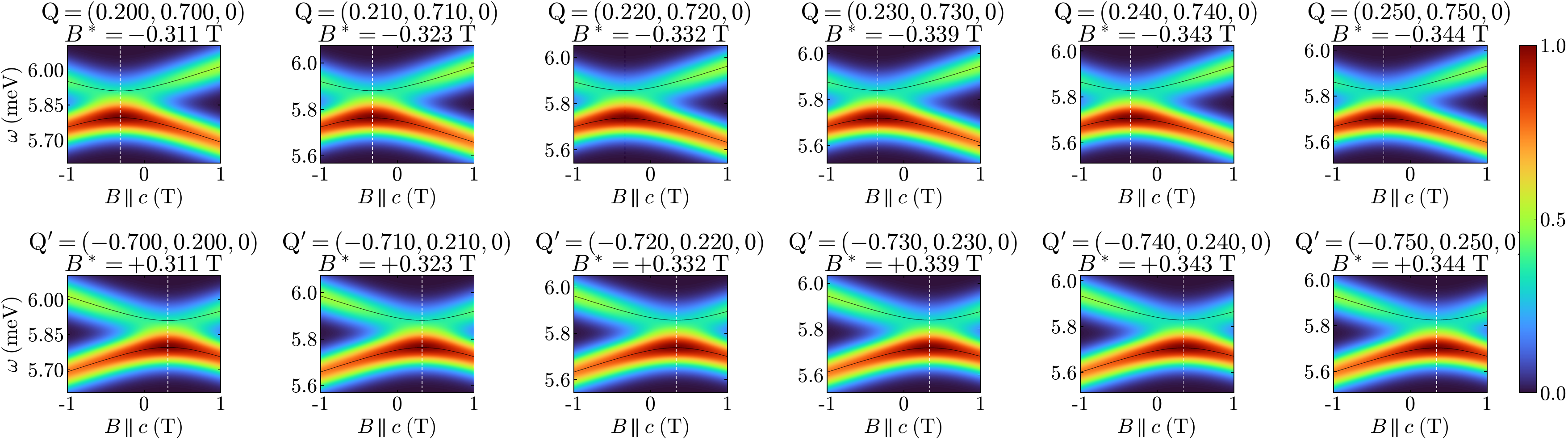}
\caption{Full BdG finite-field spectra for six representative $C_{4z}$-symmetric partner pairs within the large-splitting momentum window, defined by $\Q=(h,h+\tfrac12,0)$ and $\Q'=(-(h+\tfrac12),h,0)$. 
All cases retain substantial dipolar mixing while exhibiting pronounced zero-field doublet splittings and compensation-field shifts. The partner spectra obey $\Omega_{\Q}(B)=\Omega_{\Q'}(-B)$, where the pair at $h=0.25$ corresponds to $\Q_A,\Q_{A'}$ used in the main text. 
At $B=1~\mathrm{T}$, the doublet splittings $(\Omega_{\Q},\Omega_{\Q'})$ for the six $C_{4z}$-related momentum pairs are, from left to right: $(324.4,196.3)$, $(327.8,195.6)$, $(330.5,195.1)$, $(332.5,194.8)$, $(333.7,194.6)$, and $(334.1,194.6)~\mu\mathrm{eV}$.}
\label{fig:high-energy-partners}
\end{figure}

The first set samples the region near the maximal sublattice-odd exchange form factor:
\begin{equation}
\Q=\bigl(h,h+\tfrac12,0\bigr),\qquad
\Q'=\bigl(-(h+\tfrac12),h,0\bigr)=C_{4z}\Q.
\end{equation}
Figure~\ref{fig:high-energy-partners} displays six representative pairs within this window. Located in the higher-energy portion of the magnon band, they retain both a substantial zero-field doublet splitting and a prominent shift of the minimum away from $B=0$. The endpoint $h=0.25$ is the working pair $\mathbf{Q}_A=(0.25,0.75,0)$ and $Q_{A'}$ used in the Letter.

\begin{figure}[htbp]
\centering
\includegraphics[width=0.95\textwidth]{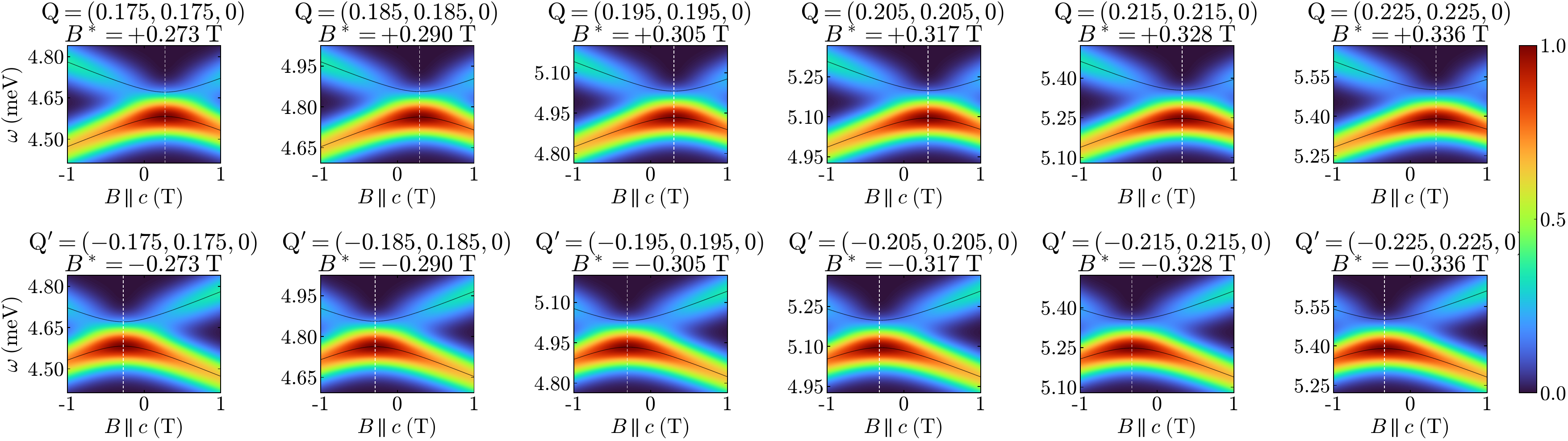}
\caption{
Full BdG finite-field spectra for six representative $C_{4z}$-symmetric partner pairs in a lower-energy momentum window, defined by $\Q=(h,h,0)$ and $\Q'=(-h,h,0)$. The upper and lower rows correspond to $\Q$ and $\Q'$, respectively, with the specific momentum labeled above each panel. Compared with the working pair used in the main text, the target doublet is shifted to lower excitation energies while retaining substantial dipolar mixing, pronounced zero-field splittings, and sizable compensation-field shifts. The opposite field polarities of each partner pair provide the same signed finite-field readout as obtained at $\Q$,$\Q'$. 
At $B=1~\mathrm{T}$, the doublet splittings $(\Omega_{\Q},\Omega_{\Q'})$ for the six $C_{4z}$-related momentum pairs are, from left to right: $(190.1,307.8)$, $(189.0,312.8)$, $(188.4,317.5)$, $(188.3,321.7)$, $(188.8,325.4)$, and $(189.8,328.6)~\mu\mathrm{eV}$.
}
\label{fig:low-energy-partners}
\end{figure}

Complementarily, a second set is chosen along
\begin{equation}
\Q=(h,h,0),\qquad
\Q'=(-h,h,0)=C_{4z}\Q,
\end{equation}
where the target doublet appears at appreciably lower excitation energies. As shown in Fig.~\ref{fig:low-energy-partners}, pronounced dipolar mixing, zero-field splitting, and compensation-field shifts persist across this entire low-energy window. Consequently, these momenta offer experimentally viable alternatives to the higher-energy working pair, without relying on an idealized dipolar-mixing-free point.

For all pairs across both sets, the full BdG spectra strictly preserve the partner relation
\begin{equation}
\Omega_{\Q}(B)=\Omega_{\Q'}(-B),
\end{equation}
such that their energy minima are shifted to equal and opposite magnetic fields. Together, these two sets establish that finite-field detection remains viable over broad regions of momentum space, providing the experimental flexibility to optimize the working point based on energy accessibility and neutron scattering intensity.

\section{Field range and stability of the collinear state}
\label{sec:s6}

The scanned range $|B|\le1$~T lies far below the spin-flop transition of \mnf{} at $9.3$~T~\cite{Jacobs1961spinflop}, and the collinear ground state of the adopted Hamiltonian was verified to remain stable over the full range at both signs of the field. The geometry $\mathbf B\parallel c$ along the easy axis induces no spin reorientation---in contrast to easy-plane MnTe, where a field rotates or selects the N\'eel vector---so the domain population is fixed during the scan and the field acts purely as a spectroscopic probe. Within this range the full BdG spectra obey the exact mirror relation of Sec.~\ref{sec:s5} at every computed field; the symmetry-allowed higher odd orders $B^3,B^5,\ldots$ in $\mathcal{C}(B)$ are numerically negligible for the production parameters ($\mathcal{C}(B)$ linear to the stated precision). For \fef{} the relevant scan range is smaller still ($|B|\lesssim0.4$~T around the $\mp0.17$~T dips) against a spin-flop scale far above it.

\section{Feasibility simulations and the global fit}
\label{sec:s7}

\emph{Synthetic experiment.} We simulate 21 field values in $[-1,1]$~T at both partner momenta---42 spectra---with Poisson counting statistics, Gaussian energy resolutions of $\Gamma_{\rm FWHM}=0.1$, $0.15$, and $0.2$~meV, and a trained $85{:}15$ domain population. The experimental inputs assumed are: neutron counts in a continuous energy window at each field and momentum; the channel widths and instrument resolution function; the momentum, field-direction, and sign conventions of Sec.~\ref{sec:s5}; the two-domain configuration and the majority-domain calibration; the staggered form-factor signs of the two momenta; and per-momentum intensity scales.

\emph{Fit parameterization.} All spectra enter one Poisson likelihood. The production fit is an effective two-mode model with free parameters
\begin{equation}
\Lambda_A,\quad D,\quad g,\quad p,\quad \omega_0,\quad \kappa,
\end{equation}
plus per-momentum signal scales and linear backgrounds; $\kappa$ is the relative spectral weight of the two mixed modes. No microscopic exchange parameter ($J_1$--$J_{12}$) is supplied to the optimizer at any stage. The optimizer determines the signed effective detuning $\Lambda_A$; only afterwards is the spin length used, as the conversion factor $\Lambda_A=4S\dJ$ of Sec.~\ref{sec:s5}. Because of the exact degeneracy $(p,\Lambda_A)\leftrightarrow(1-p,-\Lambda_A)$, the fit carries the majority-domain constraint $p>\tfrac12$ (Sec.~S5.6); $p$ may float freely or take a prior from an independent domain measurement.

\begin{figure}[htbp]
\centering
\includegraphics[width=0.85\textwidth]{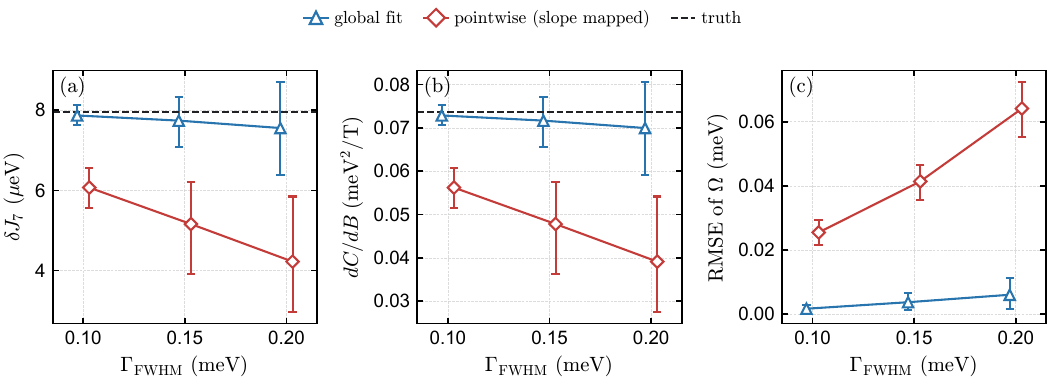}
\caption{Global versus pointwise inversion as a function of energy resolution: the global fit is unbiased at $0.1$--$0.2$~meV; pointwise fits are biased low by peak-pulling of unresolved doublets but remain sign-correct. All exchange and splitting axes use the Letter notation $\delta J_7$ and $\Omega$.}
\label{fig:s7-resolution}
\end{figure}

\begin{figure}[htbp]
\centering
\includegraphics[width=0.85\textwidth]{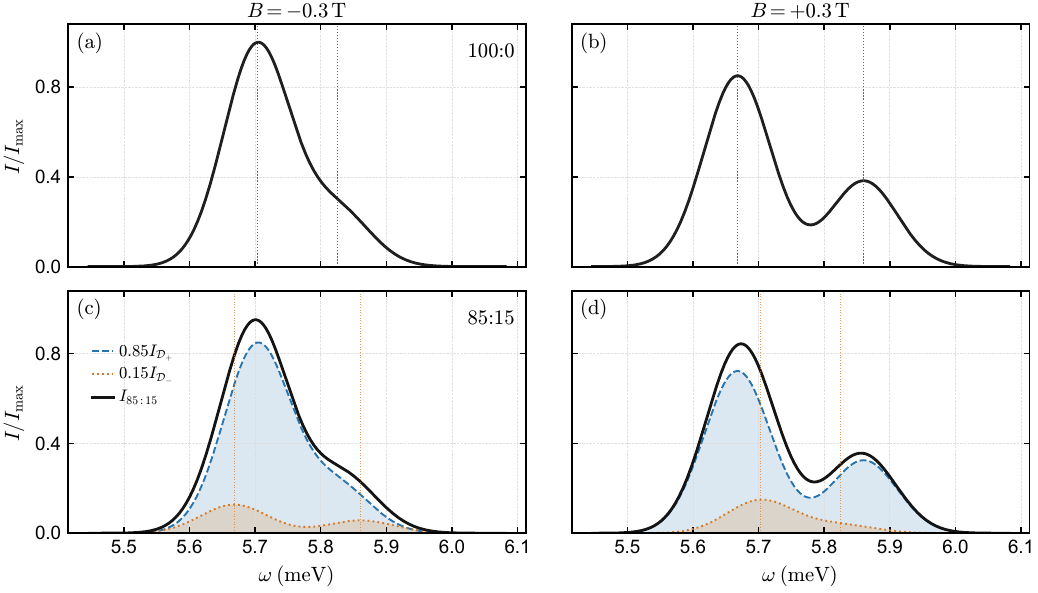}
\caption{Finite domain population ($85{:}15$): the odd-in-field contrast $\mathcal{C}(B)$ scales by $2p-1$ while the compensation-field polarity, and hence the sign readout, is unchanged.}
\label{fig:s7-domains}
\end{figure}

\begin{figure}[htbp]
\centering
\includegraphics[width=0.85\textwidth]{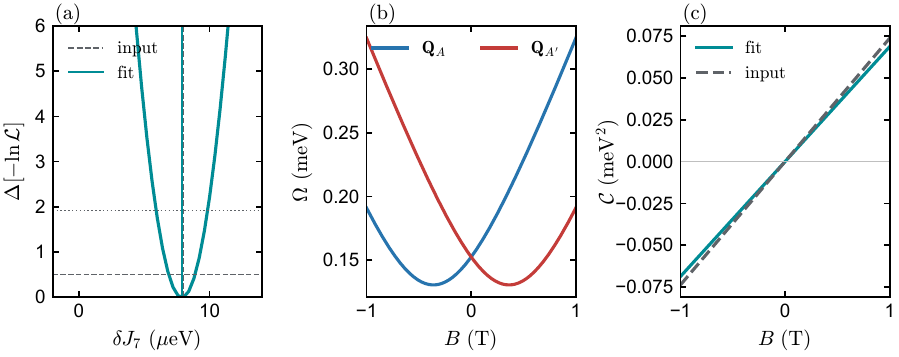}
\caption{Effective two-mode global fit on synthetic counting data: recovery of the signed detuning with all effective parameters free and no microscopic exchange input. The figure uses the Letter notation $\delta J_7$ and $\Omega$; $\Delta[-\ln\mathcal L]$ in panel (a) remains the likelihood difference.}
\label{fig:s7-globalfit}
\end{figure}

\emph{Recovery.} Against synthetic data with injected $\dJ=7.965~\mu$eV, a single fit returns $7.88~\mu$eV with 95\% profile interval $[5.91,9.84]~\mu$eV; over 100 independent Poisson replicas the mean is $7.96~\mu$eV with standard deviation $0.95~\mu$eV, and the recovered sign is correct in every replica [Fig.~\ref{fig:s7-globalfit}]. Across the three resolutions the global fit returns $\dJ=7.9/7.75/7.5~\mu$eV and the slope within 5\%---remaining nearly unbiased over the tested $0.1$--$0.2$~meV resolution range [Fig.~\ref{fig:s7-resolution}]. Deliberate model-mismatch tests (momentum-asymmetric mixing or $g$, a field-even center drift, a misstated resolution, a $10\%$ spurious peak) shift the result by less than one standard deviation. Pointwise single-field fits underestimate $|\dJ|$ by $25$--$50\%$ through peak-pulling of unresolved doublets but never miss the sign. In the $85{:}15$ domain mixture the odd-signal contrast scales by exactly $2p-1=0.70$, leaving the sign readout unchanged [Fig.~\ref{fig:s7-domains}].

\emph{Identifiability checks.} The robustness axes that enter this route directly are the energy resolution, counting statistics, domain fraction, momentum resolution, and the field-grid design; each was scanned in the simulation suite. These checks establish the conditional identifiability of the effective model at the stated counting, resolution, and field-point design; the criterion itself (Sec.~\ref{sec:s5}) is parameter-free.

\clearpage
\section{Transfer to \texorpdfstring{\fef{}}{FeF2}: parameter conversion and the two sign hypotheses}
\label{sec:s8}

The \fef{} calculations use the lattice and parameters of the published model~\cite{pins-g6dt-rf8c,M-T-Hutchings_1970}: $a=b=4.69$~\AA, $c=3.31$~\AA, $S=2$, $J_1=-0.030$~meV, $J_2=0.46$~meV, $J_4=0.005$~meV, $J_{7a}+J_{7b}=0.022$~meV, $|J_{7a}-J_{7b}|=0.005$~meV, and $0.85\times$ the free-space point-dipole strength.

\emph{Single-ion conversion.} The literature single-ion term is $-D(S^z)^2$ with $D=0.82$~meV; the present code uses the opposite sign convention together with the finite-$S$ correction of the public reference implementation, so the program input is
\begin{equation}
D_c^{\rm core}=-0.82\left(1-\frac{1}{2S}\right)=-0.615~\mathrm{meV}.
\end{equation}
With this conversion the published dispersion is reproduced; at $\Q=(0.5,0.5,1)$ the published model's dipolar splitting is $130.1977~\mu$eV against $130.1649~\mu$eV in the present implementation---a $0.033~\mu$eV difference.

\emph{Label conversion.} The literature path labels and the present geometric labels satisfy
\begin{gather}
J_{7a}^{\rm lit}=J_{7b}^{\rm here},\qquad
J_{7b}^{\rm lit}=J_{7a}^{\rm here},\nonumber\\
\text{hence}\quad
\dJ^{\rm here}=J_{7b}^{\rm here}-J_{7a}^{\rm here}=J_{7a}^{\rm lit}-J_{7b}^{\rm lit}.
\end{gather}
The published zero-field, unpolarized data constrain only $|J_{7a}-J_{7b}|$; they do not determine the sign. The two branches must therefore be written as hypotheses,
\begin{equation}
\dJ^{\rm here}=+5~\mu\mathrm{eV}
\qquad\text{or}\qquad
\dJ^{\rm here}=-5~\mu\mathrm{eV},
\end{equation}
and neither may be presented as experimentally established.

\emph{Field criterion.} For $\dJ^{\rm here}=+5~\mu$eV the numerical results are
\begin{gather}
B_A^*=-0.17276~\mathrm T,\qquad
B_{A'}^*=+0.17276~\mathrm T,\nonumber\\
\frac{d\mathcal{C}}{dB}=+0.03705~\mathrm{meV^2/T};
\end{gather}
reversing the sign of $\dJ^{\rm here}$ interchanges the two valleys and reverses the slope exactly. Both hypotheses share the identical zero-field spectrum: minimum splitting $87.008~\mu$eV and the same zero-field ordering energies. The equal-$J_7$ reference pins both valleys at $B=0$ with $\mathcal{C}(B)\equiv0$. In the unified definition of Eq.~(\ref{eq:s5-dj7}), both materials obey the same pairing,
\begin{equation}
\operatorname{sgn}\!\left(\frac{d\mathcal{C}}{dB}\right)=\operatorname{sgn}(\dJ),
\end{equation}
under the fixed $D_+$, $\Q_A/\Q_{A'}$, and field conventions of Sec.~\ref{sec:s5}.

\section{Comparison with the polarized-neutron and linear-response determinations}
\label{sec:s9}

Three independent determinations of the \mnf{} exchange imbalance exist; comparing them requires converting all to the convention of Eq.~(\ref{eq:s5-dj7}), including the $a\leftrightarrow b$ label interchange of Sec.~S5.1 and the pairs-once Hamiltonian normalization.

\emph{Polarized-neutron route.} The dispersion fit of Ref.~\cite{pins-faure2025altermagnetismrevealedpolarizedneutrons} is constrained by polarized maps but reads the imbalance from the fitted exchange model; converted to the present convention it gives
\begin{equation}
\dJ^{\rm PINS}=-4\pm4~\mu\mathrm{eV},
\end{equation}
whose $1\sigma$ interval spans zero. Two structural points temper this value. First, the zero-field dispersion entering the fit is even in $\dJ$ (Sec.~\ref{sec:s4}), so the fitted magnitude is the robust content; the sign inference is carried by the chirality maps and rests on a quantitative domain-population calibration. Second, comparing raw $\mu$eV values across parameterizations with different global scales is unreliable; the fair metric is the fractional imbalance relative to the dominant exchange, $\dJ/J_2$: $+2.35\%$ for the present mapping against $-1.39\%$ for the converted fit. The magnitudes are compatible at the one-standard-deviation level; the central signs are opposite. The conversion itself is the $a\leftrightarrow b$ relabel of Sec.~S5.1 together with the pairs-once Hamiltonian normalization already used throughout: no further worksheet is required. The Letter's primary claim is the compensation-field displacement $B^\ast_A=-B^\ast_{A'}$, not an adjudication of these three central values.

\emph{Linear-response route.} The independent linear-response calculation of Ref.~\cite{mnf-band-solovyev2026altermagnetismmnf2bandsplitting} resolves the same symmetry-split pair (there labeled $J_4\pm\delta J_4$; the $[110]$ fluorine-bridged path is $J_4+\delta J_4$). Its Hamiltonian pairs each bond once with no hidden normalization, so the conversion to the present convention is a sign bookkeeping only, giving $\dJ=+35.8~\mu$eV with the same $\sim$20--25:1 path hierarchy. The linear-response route thus independently supports the positive sign and the path hierarchy of the present mapping.

\begin{table}[htbp]
\caption{Three existing determinations of the \mnf{} exchange imbalance, converted to the convention $\dJ=J_{7b}-J_{7a}$ of Sec.~\ref{sec:s5}. The Letter reports $|\dJ|$ and the opposite partner shifts $B^\ast_A=-B^\ast_{A'}$ rather than using the field scan to choose among these central values.}
\label{tab:s9-compare}
\centering
\begin{tabular}{llll}
\toprule
Route & $\dJ$ (converted) & Fractional $\dJ/J_2$ & Sign basis \\
\midrule
Total-energy mapping (this work) & $+7.96~\mu$eV ($+5.4$ to $+11.4$) & $+2.35\%$ & robust: all $U_{\mathrm{eff}}$, $10^4$ replicas \\
Linear response~\cite{mnf-band-solovyev2026altermagnetismmnf2bandsplitting} & $+35.8~\mu$eV (scale $\approx3\times$ hot) & $+3.5\%$ & same sign, same path hierarchy \\
PINS-constrained fit~\cite{pins-faure2025altermagnetismrevealedpolarizedneutrons} & $-4\pm4~\mu$eV & $-1.39\%$ & via domain calibration; $1\sigma$ spans 0 \\
\bottomrule
\end{tabular}
\end{table}

\clearpage
\bibliography{mnf2-cite}